\documentstyle[aps,prb]{revtex} 
\input epsf.sty 

\begin{document}
\renewcommand{\theequation}{\arabic{section}.\arabic{equation}} 

\twocolumn[ 
\hsize\textwidth\columnwidth\hsize\csname@twocolumnfalse\endcsname 
\draft 

%%%%%%%%%%%%%%%%%%%%%%%%%%%%%%%%%%%%%%%%%%%%%%%%
\title{CP$^1$+ U(1) Lattice Gauge Theory in Three Dimensions:\\
Phase Structure, Spins, Gauge Bosons, and Instantons} 
\author{Shunsuke Takashima and Ikuo Ichinose} 
\address{Department of Applied Physics,
Nagoya Institute of Technology, Nagoya, 466-8555 Japan 
} 
\author{Tetsuo Matsui} 
\address{Department of Physics, Kinki University, 
Higashi-Osaka, 577-8502 Japan 
} 
\date{\today}

\maketitle 

\begin{abstract}   
In this paper we study a 3D lattice spin model of CP$^1$ Schwinger-bosons 
coupled with dynamical compact U(1) gauge bosons.
The model contains two parameters; the gauge coupling and 
the hopping parameter of CP$^1$ bosons.
At large (weak) gauge couplings, the model reduces to 
the classical O(3) (O(4)) spin model with 
long-range and/or multi-spin interactions.
It is also closely related to the recently proposed 
``Ginzburg-Landau" theory for quantum phase transitions of 
$s=1/2$ quantum spin systems on a 2D square lattice at zero temperature.
We numerically study the phase structure of the model 
by calculating specific heat,
spin correlations, instanton density, and gauge-boson mass.
The model has two phases separated by a critical line of second-order 
phase transition; O(3) spin-ordered phase and spin-disordered phase.
The spin-ordered phase is the Higgs phase of 
U(1) gauge dynamics, 
whereas the disordered phase is the confinement phase.
We find a crossover in the confinement phase
which separates dense and dilute regions of instantons.
On the critical line, spin excitations are gapless, but
the gauge-boson mass is {\it nonvanishing}. This
indicates that a confinement phase is realized 
on the critical line. To confirm this point, we also study the
noncompact version of the model. 
A possible realization of a deconfinement phase 
on the criticality is discussed for
the CP$^N$+U(1) model with larger $N$.
\end{abstract} 
]%the end of twocolumn 

\setcounter{footnote}{0} 
%%%%%%%%%%%%%%%%%%%%%%%%%%%%%%%%%%%%%%%%%%%%%%%%%%%%%%%%%%%%%%%%%

\section{Introduction}
\setcounter{equation}{0}

The concept of gauge theory plays an important role not only in 
elementary particle physics but also in condensed matter physics.
In particular, in strongly-correlated electron systems and 
quantum spin systems at quantum
phase transitions, unconventional 
low-energy excitations with ``fractional" and/or ``exotic" 
quantum numbers are expected to appear.
In order to describe these excitations, 
gauge-theoretical approaches are sometimes quite useful.

For example, in the fractional quantum Hall 
effect (FQHE),  each electron in a strong magnetic field
separates into a so-called composite fermion
(CF) and even number of fictitious magnetic flux quanta.\cite{cf} 
Various experiments indicate that 
the CF's, instead of electrons, behave as quasiparticles.
In a gauge theoretical approach to describing 
this phenomenon, a gauge field binding each CF 
and flux quanta is introduced\cite{pfs}. 
When the dynamics of this gauge field 
is realized in a {\em deconfinement} phase, CF's and flux quanta
are to dissociate each other, and the CF's become quasiparticle
moving in an effective magnetic field made of the external magnetic
field and the condensation of flux quanta.
Then the integer QHE of CF's explains the FQHE.
This mechanism of FQHE was called particle-flux separation (PFS),
and the critical temperature of confinement-deconfinement
transition below which the CF picture holds
was estimated\cite{pfs}.  

In the high-T$_{\rm c}$ superconductivity,
a phenomenon called charge-spin separation (CSS)\cite{anderson}
was proposed to explain the anomalous behavior of the metallic
phase of cuprates.
In the CSS scenario, each electron in a cuprate is viewed as
a composite of a holon  and a spinon.
A gauge-theoretical understanding
of the CSS parallels the PFS as  a deconfinement
phenomenon of the gauge field binding a holon and a 
spinon\cite{css1,css2}. 
Then holons and spinons themselves behave as quasi-free particles. 

Recently, importance of quantum phase transitions (QPT) has
been recognized, and several interesting proposals have been 
made\cite{qpt}.
One of the most tractable systems of QPT is quantum spin
models in low dimensions.
By varying some parameters of a quantum spin Hamiltonian,  
a phase transition from one phase with a certain order (e.g.
the N\'eel order) to another phase with different order (e.g.
the dimer) may occur.
Then one is interested in what is 
the low-energy effective theory of the system
that corresponds to the Ginzburg-Landau
theory (GLT) of a second-order phase transition,
and how it describes the phase transition. 
One may also ask the nature of the low-energy excitations 
appearing just on the criticality.

For the conventional superconductivity, the GLT is well known 
and involves a collective scalar field describing Cooper pairs 
and a gauge field describing electromagnetic (EM) vector potential.
The superconducting phase corresponds to the Higgs phase of
the EM gauge dynamics, in which  gauge bosons (photons) acquire a 
mass via the Anderson-Higgs mechanism. The normal phase correponds 
to the Coulomb phase with massless gauge bosons. 

For a certain class of quantum $s=1/2$ spin models in two 
dimensions, it  has been proposed that the correponding ``GLT" is
 a CP$^1$ model coupled with a dynamical
U(1) gauge field\cite{fisher}. This is quite natural because
the path-integral representation involves
a set of coherent states of $s=1/2$ quantum spin, which is
conveniently labelled  by  a CP$^1$ 
(complex projective) variable, sometimes called Schwinger boson.
The U(1) gauge field here appears as  $s=0$ collective
excitations to generate effective interactions between CP$^1$ spinons.

Then it is an urgent matter to set up a concrete CP$^1$ model
coupled with a U(1) gauge field, and study its phase structure,
low-energy excitations, etc.
In this paper, we shall consider a three-dimensional (3D) 
lattice CP$^1$ model and study its properties by numerical methods
very intensively. 
The model contains two parameters $c_1$ and $c_2$; $c_1$ 
controls the spin stiffness (the hopping parameter of CP$^1$ bosons) 
and $c_2$ is the inverse gauge coupling constant which controls 
fluctuations of the U(1) gauge field.
This model may be viewed as 
an effective model of a quantum spin system on a 2D square lattice
at zero temperature ($T$). 
It is also closely related to the t-J model of the cuprates\cite{css1,css2} 
and the gauge-theory approach
to heavy fermion systems\cite{hf}.

Let us briefly explain the relation between the present model
and the t-J model of strongly correlated electrons.
In the previous papers\cite{sf}, we studied
the t-J model in the slave-fermion (SF) representation,
in which spinons are described by CP$^1$ bosons and
holons are described by fermions.
At the half filling (no holons), the system becomes 
the 2D  $s=1/2$ antiferromagnetic (AF) Heisenberg spin model.
At $T=0$ it is described by the 3D CP$^1$ model with infinite
gauge coupling, which is just the present model at $c_2=0$.
The model exhibits a phase transition at $c_1= c_{1c}$
between the  O(3) spin-ordered and disordered phases\cite{css1}. 
(See Sec.2B and Sec.4 for more details.)
In lightly doped cases, holons can be integrated to
``renormalize"  the spin-stiffness constant $c_1$ 
and destroy the long-range N\'eel order 
as the holon density increases\cite{sf}. 
For larger dopings, holons
generates nontrivial interactions between the gauge bosons.
Such a system may be mimicked to certain extent
by the present model with a suitable amount of the $c_2 $-term
[See Sec.2B(ii)], 
although one needs certainly more quantitative study on this point.

The rest of the paper is organized as follows.
In Sec.2, we shall introduce the model and explain its relations
to the classical O(3) and O(4) nonlinear sigma models, 
the GLT for quantum spin models, etc.
In Sec.3, we study the phase diagram  by mean-field theory (MFT).
In Sec.4, we calculate the specific heat and present the global phase
diagram of the model in the $c_2-c_1$ plane. 
The model exhibits a second-order phase transition 
along the critical line $c_1 = c_{1c}(c_2)$, which separates
confinement phase and Higgs phase.
In Sec.5, we study spin-spin correlation functions.
The results shows that the observed phase transition accompanies
a spontaneous magnetization and the spin excitations are gapless 
on the critical line as well as in the spin-ordered phase. 
In Sec.6, we calculate instanton (monopole) density in various
places in the phase diagram.
This supplies us with useful information to understand the gauge dynamics.
In Sec.7, we measure the gauge-boson mass by calculating 
gauge-invarinat correlation functions of field strength.
Two mechanisms of generating a gauge-boson mass are explained;
(i) condensation of instantons and (ii) Anderson-Higgs mechanism
by condensation of the CP$^1$ variables.
We also obtain the gauge-boson mass in $\lambda \phi^4$ Higgs
models and compare the results with the previous calculations
in order to check the present calculations.
Section 8 is devoted to discussion.

%%%%%%%%%%%%%%%%%%%%%%%%%%%%%%%%%%%%%%%%%%%%%%%%%%%%%%%%%%%%%%%%%%%%%%%%%%%%%%

\section{The 3D CP$^1$+U(1) lattice gauge model}
\setcounter{equation}{0}
\subsection{Model}

Let us define the CP$^1$+U(1) gauge model on the cubic lattice.
We use $x$ as the site index, and $\mu=0,1,2$ as the direction index.
We use $\mu$ also as the unit vector in the $\mu$-th direction. 
On each site $x$ we have a CP$^1$ variable $z_x$, which is a 
two-component complex number,
\begin{equation}
z_x\equiv \left(\begin{array}{c}
z_{x1}\\
z_{x2}\end{array}
\right), \,\,\,\,\,\,\,\,\,\, z_{x1},z_{x2}\in C,
\end{equation}
satisfying the CP$^1$ constraint,
\begin{equation}
 \sum_{a=1}^2 |z_{xa}|^2=1.
\label{CP1con}
\end{equation}
On each link $(x,x+\mu)$ we have a U(1) gauge variable,
$U_{x\mu}=\exp(i\theta_{x\mu})$ [$\theta_{x\mu}\in (-\pi,+\pi)$].

The partition function $Z$ of the model is written as
\begin{eqnarray}
Z&=&\int [dU]\int[dz] \exp(-S),\nonumber\\
\left[dU\right] &\equiv& \prod_{x,\mu}dU_{x\mu}=\prod_{x,\mu}
\frac{d\theta_{x\mu}}{2\pi},\nonumber\\
\left[dz\right]
&\equiv& \prod_{x}dz_{x1}dz_{x2}\delta(|z_{x1}|^2
 +|z_{x2}|^2-1).
\label{Z}
\end{eqnarray}
The action $S$ of the model is given by
\begin{eqnarray}
S&=&-\frac{c_1}{2}\sum_{x,\mu,a}\Big(\bar{z}_{x+\mu,a}U_{x\mu} z_{xa} + 
\mbox{H.c.}\Big) \nonumber  \\
&&
-\frac{c_2}{2}\sum_{x,\mu<\nu}\Big(\bar{U}_{x\nu}\bar{U}_{x+\nu,\mu}
U_{x+\mu,\nu}U_{x\mu}+\mbox{H.c.}\Big),
\label{model_1}
\end{eqnarray}
where $c_1$ and $c_2$ are real parameters of the model.

The action $S$ is invariant under the following U(1) local
($x$-dependent) gauge transformation;
\begin{eqnarray}
z_{xa}&\rightarrow& z'_{xa}= \exp(i\Lambda_x)z_{xa},\nonumber\\
U_{x\mu}&\rightarrow& U'_{x\mu}= \exp(i\Lambda_{x+\mu})
U_{x\mu}\exp(-i\Lambda_x).
\label{gaugesymmetry}
\end{eqnarray}
A genuine CP$^1$ variable is characterized by (i) the constraint 
(\ref{CP1con}) and (ii) the redundancy under the phase transformation
$z_{xa} \rightarrow \exp(i\Lambda_x) z_{xa}$. The gauge symmetry  
(\ref{gaugesymmetry}) assures us of this redundancy.

Besides the above local gauge symmetry, $S$ has a symmetry 
under the following global SU(2) transformation,
\begin{equation}
z_x \rightarrow V z_x, V \in {\rm SU(2)}.
\label{SRS}
\end{equation}
This induces the global O(3) spin rotation, which
shall be explained in the following subsection B(ii).

%%%%%%%%%%%%%%%%%%%%%%%%%%%%%%%%%%%%%%%%%%%%%%%%%%

\subsection{Special cases}

The present model reduces to some known models in certain limits of the 
parameters.\\

(i) Pure gauge model ($c_1=0$)\\
For $c_1=0$ the model reduces to the 3D compact U(1) gauge model without
matter couplings, where $c_2$ is the inverse of the gauge coupling constant
$g$, i.e., $c_2 \propto g^{-2}$.
Polyakov\cite{polyakov} showed that this model
stays always in the confinement phase regardless of the value of $c_2$
due to the condensation of instantons (monopoles).
We shall study the instanton effect and verify his result 
by numerical calculation in Sec.6.\\

(ii) O(3) spin model ($c_2=0$)\\
For $c_2=0$ the model reduces to a O(3) nonlinear sigma model
(classical Heisenberg spin model) with a modified action.
Here one can perform the integration over $U_{x\mu}$ exactly because
the plaquette coupling between $U_{x\mu}$'s disappears. 
One obtains
\begin{equation}
Z=\int [d\vec{S}] \exp\Big[\sum_{x,\mu}\ln I_0\Big(c_1
\sqrt{\frac{1+\vec{S}_{x+\mu}
\cdot \vec{S}_{x}}{2}}\Big)\Big],
\label{I0}
\end{equation}
with the following measure,
\begin{equation}
 [d\vec{S}] = \prod_{x}\prod_{i=1}^3 \delta(\sum_{i=1}^3
S_{xi}^2 -1),
\label{measure}
\end{equation}
where $I_0(\gamma)$ is the modified Bessel function.
$\vec{S}_x$ is a three-component O(3) clasical spin vector,
made of the CP$^1$ variable,
\begin{eqnarray}
\vec{S}_x=\bar{z}_x\vec{\sigma} z_x,\ \ \  \vec{S}_x \cdot \vec{S}_x=1,
\label{o3spin}
\end{eqnarray}
where $\vec{\sigma}$ is the $2\times2$ Pauli spin matrices.
We note that $\vec{S}_x$ is a gauge-invariant object.
A global O(3) rotation  of $\vec{S}_x$, which is a symmetry of the model,
is induced by a global SU(2) transformation of $z_x$,
\begin{eqnarray}
\vec{S}_x &\rightarrow& \Omega \vec{S}_x,
\Omega \in {\rm O(3)}, \nonumber\\  
z_x &\rightarrow& V z_x, V \in {\rm SU(2)}. 
\label{su2spin}
\end{eqnarray}
In Eq.(\ref{o3spin}) we have used the relation 
$[dz] f(\{\bar{z}_x\vec{\sigma} z_x\})=
[d\vec{S}]f(\{\vec{S}_x\})$.
Since $\ln I_0(\gamma)$ is a monotonically increasing function 
of $\gamma$,
it is almost obvious that the model (\ref{I0}) belongs to the same 
universality class as the standard
O(3) model with the action $-J\sum \vec{S}_{x+\mu}\cdot\vec{S}_x$.
Then the model (\ref{I0}) should exhibit a second-order phase transition
at a certain critical value $c_1=c_{1c}({\rm O(3)})$, which
separates the O(3) spin-ordered phase and the disordered phase.
It is obvious from Eq.(\ref{I0}) that the parameter $c_1$ controls 
the spin stiffness.

For small $c_2$, integration over the gauge field $U_{x\mu}$ can be done
perturbatively in powers of $c_2$.
The plaquette term in Eq.(\ref{model_1}) generates four-spin couplings
like $(\vec{S}_{x+\mu}\cdot \vec{S}_{x})
(\vec{S}_{x+\mu+\nu}\cdot \vec{S}_{x+\nu})$ in the leading order of $c_2$, 
which prefer a ferromagnetic order for $c_2 > 0$.
The higher-order terms of $c_2$ become nonlocal and contain 
many spin variables.

For any finite value of $c_2$, integration over the gauge field 
$U_{x\mu}$ in the partition function $Z$ in (\ref{Z}) results in an
effective spin model $S_{\rm{SM}}(\vec{S}_{x})$,
\begin{equation}
\int [d\vec{S}]\exp (-S_{\rm{SM}})=\int [dU]\int[dz] \exp(-S),
\label{ZSM}
\end{equation}
because ``spin variable" $\vec{S}_{x}$ is the only gauge-invariant 
object made of $z_{xa}$.
In later discussions, we call the off-diagonal long-range order of 
$\vec{S}_{x}$ magnetization since it measures a spontaneous breakdown of the
SU(2) or O(3) global symmetry (\ref{su2spin}).
Later numerical calculation shows that there exists a second-order 
phase transition in the present model for which $\vec{S}_{x}$ works
as an order parameter.

(iii) O(4) spin model ($c_2= \infty$)\\
In the limit $c_2\rightarrow \infty$, fluctuations of $U_{x\mu}$ are 
totally suppressed and $U_{x,\mu}$ is restricted to pure gauge
configurations, i.e., $U_{x\mu} = \exp(-i\Lambda_{x+\mu})\exp(i\Lambda_x)$. 
Then $Z$ is reexpressed in terms of four-component 
O(4) classical spin variables $R_{x\alpha} (\alpha=1,2,3,4) \in R$,
\begin{eqnarray}
z_{x1}&=& R_{x1}+iR_{x2}, \ \ z_{x2}= R_{x3}+iR_{x4}, \nonumber\\
\bar{z}_x z_x &=& \sum_{\alpha=1}^4 R_{x\alpha}^2 =1,\nonumber\\
\left[dz\right] 
&=& [d\vec{R}] = \prod_x \prod_{\alpha=1}^4 dR_{x\alpha}\cdot
\delta(\sum_{\alpha}R_{x\alpha}^2-1)
\end{eqnarray}
as 
\begin{eqnarray}
Z &=& \int [dR] \exp(c_1\sum_{x,\mu}\vec{R}_{x+\mu}\vec{R}_x).
\end{eqnarray}
This model is just the 3D nonlinear O(4) sigma model which 
exibits also a second-order phase transition at 
$c_{1}=c_{1c}({\rm O(4)})$,
which separates the O(4) spin-ordered phase and the disordered phase.\\

%%%%%%%%%%%%%%%%%%%%%

\subsection{The nonuniform AF Heisenberg model}

\begin{figure}[bthp]
\begin{center}
\leavevmode
\epsfxsize=8cm
\epsffile{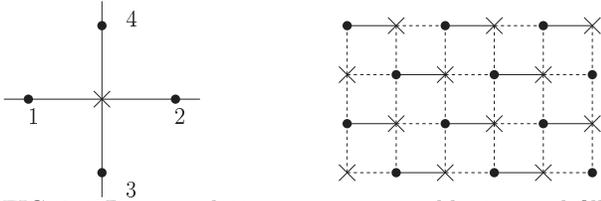}
\caption{2D square lattice; crosses are odd sites, and filled
circles are even sites. Solid bonds indicate that their
exchange couplings are stronger than those on dotted bonds.}

\label{fig.dimer}
\end{center}
\end{figure}

Let us explain how the model Eq.(\ref{model_1}) is related to the recently 
proposed effective field theories (GLT's) for the {\em quantum magnets}.
In Ref.\cite{YAIM}, we studied a phase transition from the 
N\'eel state to the dimer state in the nonuniform $s={1\over 2}$
AF Heisenberg model on a square lattice,
\begin{equation}
H_{\rm AF}=\sum_{x,j}J_{xj}\vec{\hat{S}}_x\cdot\vec{\hat{S}}_{x+j},
\label{AFH}
\end{equation}
where $j$ is the spatial direction
index ($j=1,2$), $\vec{\hat{S}}_x$ is the quantum spin operator at site $x$
and $J_{xj}$ is the nonuniform exchange coupling.
We rename the even lattice sites $x=(o,i)$ where $o$ denotes each odd site,
and the index $i=1,2,3$ and $4$ specifies its four nearest-neighbor(NN)
even sites (see Fig.\ref{fig.dimer}).

%---------------------------------------------------
%----------------------------------------------------

The exchange couplings $J_{xj}=J_{oi}$ may be position dependent,
and in Ref.\cite{YAIM} we explicitly considered the following 
dimer configuration,
\begin{eqnarray}
J_{oi}&=&J+\Delta J_{oi},   \nonumber  \\
\Delta J_{oi}&=&\left\{\begin{array}{ll}
              \Delta J_{oi}=\alpha J, & i=1 \\
              \Delta J_{oi}=-\alpha J, & i=2,3,4
                      \end{array}
    \right.
\label{J}
\end{eqnarray}
where $0\le\alpha\le 1$ is a parameter which interpolates 
between the uniform
Heisenberg model ($\alpha=0$) and the dimer model ($\alpha=1$).
Its  effective field theory is derived 
by the standard method\cite{sf,aa} in path-integral formalism by 
integrating over the odd-site spins. It takes a form
of a CP$^1$ model in (2+1)-dimensions (the extra one dimension
is the direction along the imaginary time).
Namely, the effective theory has the form of Eq.(\ref{model_1}), 
where the effective parameter $c_1$ is expressed
in terms of $\alpha$ and the lattice spacing $a$ as  
\begin{equation}
c_1 ={1\over 2\sqrt{2}a}\cdot
{1-\alpha\over 2-\alpha}\sqrt{2(2+\alpha)\over 1-\alpha}.
\label{fb} 
\end{equation}
The bare effective parameter  $c_2$ is vanishing in this effective
field theory.

As explained in the case (ii) above, the model with $c_2=0$ exhibits
a second-order phase transition at $c_1 = c_{1c}({\rm O(3)})$.
Through the correspondence (\ref{fb}), this transition
from the ordered state to the disordered state of the CP$^1$ 
model describes
the transition of the nonuniform AF magnet from the N\'eel
state to the dimer state\cite{ND}.
In Ref.\cite{YAIM} we pointed out the possibility that 
the {\em spinons} $z_{xa} \; (a=1,2)$ appear as gapless
excitations {\em at the critical point}.
That is, a deconfinement phase like the Coulomb phase in the
(3+1)-dimensions is realized there because of the appearance 
of long-range correlations among the gauge field.
In this paper, among others, we shall reexamine this point,
i.e., investigate the possibility whether  
the Coulomb phase is realized on the critical line of 
the model (\ref{model_1}).

%%%%%%%%%%%%%%%%%%%%%%%%%%%%%%%

\subsection{The uniform AF Heisenberg model with ring exchange}

\begin{figure}[htbp]
\begin{center}
\leavevmode
\epsfxsize=8cm
\epsffile{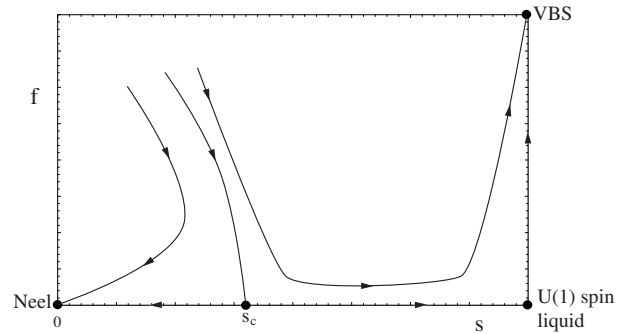}
\caption{Renormalization-group flows of the effective gauge theory
of the AF Heisenberg model with the ring exchange\cite{fisher}.
The horizontal axis refers to the parameter $s$,
and the vertical line to the instanton fugacity $f$.}
\label{fig.flow}
\end{center}
\end{figure}

In a framework of the uniform AF Heisenberg model with multiple
spin ring exchange, Senthil et al.\cite{fisher} argued that a phase 
transition from the
N\'eel state to valence bond solid (VBS) takes place.
They suppose that the system in the N\'eel state but near the  
{\em critical point} is described by a CP$^1$ 
field coupled 
to a {\em noncompact} U(1) gauge field, where nonperturbative
contributions from
instantons are totally suppressed.
An effective field theory 
in the continuum space-time in this region was given by\cite{fisher}
\begin{eqnarray}
{\cal L}_{\mbox{\footnotesize{deconfine}}}
&=& |(\partial_\mu -iA_\mu)\tilde{z}_a|^2 +s |\tilde{z}_a|^2 +
u(|\tilde{z}_a|^2)^2  \nonumber  \\
&& +{1\over g^2}(\epsilon_{\mu\nu\lambda}\partial_\nu A_\lambda
)^2,
\label{Ldec}
\end{eqnarray}
where $\tilde{z}_a \; (a=1,2)$ are {\it unconstrained} 
complex scalar fields coming from
the CP$^1$ spin variables, $A_\mu$ is the gauge potential, and
$s$ and $u$ are parameters.
For $s<s_c$,  $\tilde{z}_a$ are condensed 
$\langle \tilde{z}_a \rangle \neq 0$
which corresponds to the long-range N\'eel order.
The gauge dynamics is in the Higgs phase and the gauge bosons
acquire a  mass gap $|\langle \tilde{z} \rangle |$.
At the criticality $s=s_c$, $\langle \tilde{z}_a \rangle = 0$,
whereas the effect of instantons is negligible because of the 
cancellation mechanism via the Berry phase term\cite{Bphase}.
On the other hand, in the VBS state, the instatons are relevant and 
proliferate to generate the confinement phase.
The proposed phase diagram\cite{fisher} 
of the AF Heisenberg model is shown in
Fig.\ref{fig.flow}. We shall compare it  with that of the present 
model in Sec.7C.

%---------------------------------------------------
%----------------------------------------------------

%%%%%%%%%%%%%%%%%%%%%%%%%%%%%%%%%%%%%%%%%%%%%%%%%%%%%%%%%%%%%%%%%%%%%%

\section{Mean field theory}
\setcounter{equation}{0}

Before going into detailed numerical studies of the model, 
let us apply a MFT  to the model
to obtain a rough phase diagram.

The MFT below is based  upon the variational principle\cite{feynman}.
We start with the variational action $S_0$ and use the following
relations;
\begin{eqnarray}
Z_0& \equiv& \int [dz]  [dU] 
\exp(-S_0) \equiv \exp(-F_0),\nonumber\\
F &\equiv& -\ln Z \le  F_v \equiv F_0 + \langle S -S_0 \rangle_0, 
\nonumber\\
\langle O \rangle_0 &\equiv& Z_0^{-1} 
\int [dz]  [dU] \ O\ \exp(-S_0).
\end{eqnarray}
From this Jensen-Peierls inequality, we adjust the variational parameters
contained in $S_0$ optimally so that $F_v$ is minimized.

For the trial action $S_0$,
we assume the translational invariance
and consider the following sum of single-site and single-link energies;
\begin{eqnarray}
S_0 = - \frac{1}{2} \sum_{x}(\sum_{\mu}\bar{W}U_{x\mu}
+\sum_{a}\bar{H}_{a} z_{xa} + c.c.),
\end{eqnarray}
where $W$ and $H$ are complex variational parameters. 
Then we obtain the following free energy per site, $f_v \equiv F_v/N$,
where $N$ is the total number of the 3D lattice sites 
(we present the formulae for $d$-dimensional lattice);  
\begin{eqnarray}
h &\equiv& |H|,\ \omega \equiv |W|, \nonumber\\ 
f_v &=& -d \ln (I_0(\omega))
-\ln(I_0(h) -I_2(h))
 -c_1 dm^2 p  \nonumber \\
& -&c_2 \frac{d(d-1)}{2} p^4 + d\omega p + hm,\nonumber\\
m &\equiv& (\sum_a |\langle z_{xa} 	\rangle_0|^2)^{1/2} =  
\frac{1}{2}\frac{I_1(h)-I_3(h)}
{I_0(h)-I_2(h)},\ \  \nonumber\\
p &\equiv& |\langle U_{x\mu} 	\rangle_0| = 
\frac{I_1(\omega)}{I_0(\omega)},
\label{fv}
\end{eqnarray}
where $I_n(\gamma)$ ($n$ is an integer) is the modified Bessel function,
\begin{eqnarray}
I_{n}(\gamma) &=& \int_{0}^{2\pi} \frac{d\theta}{2\pi}
\exp(\gamma \cos\theta + i n \theta).
\end{eqnarray}
The stationary conditions for $f_v$ w.r.t. $\omega, h$ read
\begin{eqnarray}
\omega &=& c_1 m^2 +2c_2 (d-1)p^3 +6 c_3 (d-1)m^2 p^2, \nonumber\\
h &=& 2dc_1 mp +4c_3 d(d-1)mp^3.
\label{stationary}
\end{eqnarray}

The MFT equations (\ref{fv})-(\ref{stationary}) for $d=3$ predict
the existence of 
the three phases characterized as follows;
\begin{eqnarray}
\begin{tabular}{|c|c|c|} 
\hline
   phase    & $p=\langle U_{x\mu} \rangle $ & $m=\langle z_x \rangle $  
\\ \hline
Higgs       & $\neq 0$  & $\neq 0$  \\ \hline
Coulomb     & $ \neq 0$  & $0$  \\ \hline
Confinement & $0$   & $0$     \\
\hline
\end{tabular}
\end{eqnarray}
The naming of each phase should be clear.
We note that the fourth combination $p = 0$ and $m \neq 0$ is missing.
%---------------------------------------------------
\begin{figure}[htbp]
\begin{center}
\leavevmode
\epsfxsize=8cm
\epsffile{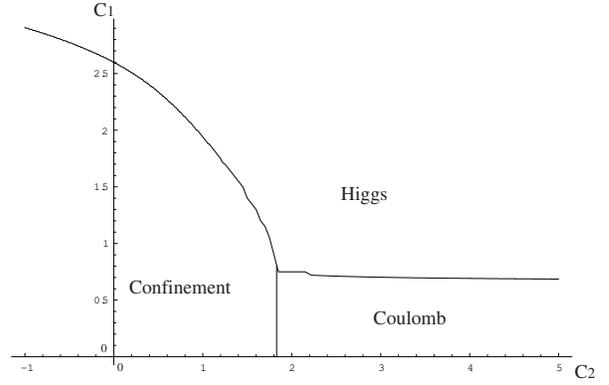}
%\epsffile{3dim_cp1_MFT_0331_2.eps}
\caption{Phase diagram of the 3D CP$^1$+U(1) model in MFT.
There are three (Higgs, confinement, Coulomb) phases.
}
\label{fig.mft}
\end{center}
\end{figure}
%----------------------------------------------------
In Fig.\ref{fig.mft}, we plot the phase boundaries 
obtained from Eqs.(\ref{fv})-(\ref{stationary}).
The phase boundary of the MFT between the Higgs and Coulomb
phases is of second order, 
whereas other two boundaries, the Higgs-confinement 
and confinement-Coulomb, are of first order.
Across the second-order transition, $p$ and $m$ vary continuously,
while across the first-order transitions, $p$ and/or $m$ change 
discontinuously with finite jumps of $\Delta p$ and/or
$\Delta m$. For the Higgs-confinement transition,
$\Delta p \neq 0$ and $\Delta m \neq 0$, and
for the confinement-Coulomb transition,
$\Delta p \neq 0$ and  $\Delta m = 0$ since $m = 0$ in both phases.

The present MFT for the CP$^1+U(1)$ model may 
have some inappropriate points:

(i) As explained above, for the pure gauge case $c_1 = 0$,
the confinement phase should survive up to $c_2 \rightarrow \infty$ 
and the Coulomb phase in the MFT should disappear.
In fact we verify the nonexistence of the Coulomb phase 
in the following section.
However we shall see some crossover phenomenon
close to this ficticious transition line. 
See later discussion.

(ii) At $c_2=0$, the system reduces to the O(3)
spin model (\ref{I0}) having a second-order transition at $c_1=c_{1c}
({\rm O(3)})$. 
Thus the first-order transition of the MFT at $c_2=0$ is incorrect. 
This MFT result should be modified by Monte Carlo simulations
in the following section.

On the other hand, in the limit $c_2 \rightarrow \infty$, the
MFT correctly predicts the order of transition, i.e.,
the second order one of the O(4) spin model.

%%%%%%%%%%%%%%%%%%%%%%%%%%%%%%%%%%%%%%%%%%%%%%%%%%%%%%%%%%%%%%%%%%%%%%

\section{Phase structure}
\setcounter{equation}{0}

To study the phase structure of the model numerically,
we performed Monte-Carlo simulations with Metropolis algorithm
for the 3D cubic lattice of the size up to $N=L^3=30^3$
with the periodic boundary condition (PBC).

\begin{figure}[htbp]
\begin{center}
\leavevmode
\epsfxsize=8.5cm
\epsffile{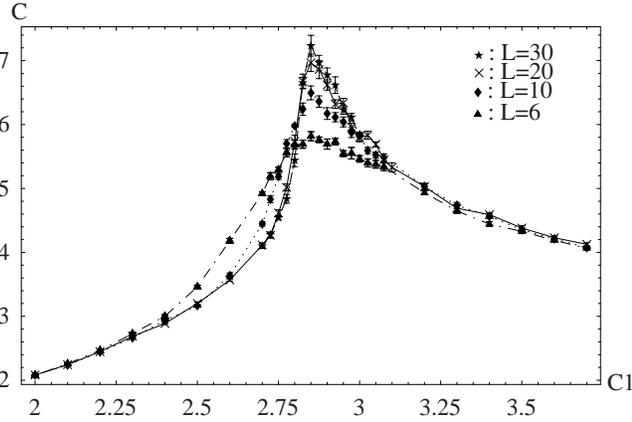}
\caption{System-size dependence of the specific heat for $c_2=0$.}
\label{fig.sh1}
\end{center}
\end{figure}

\begin{figure}[htbp]
\begin{center}
\leavevmode
\epsfxsize=8.5cm
\epsffile{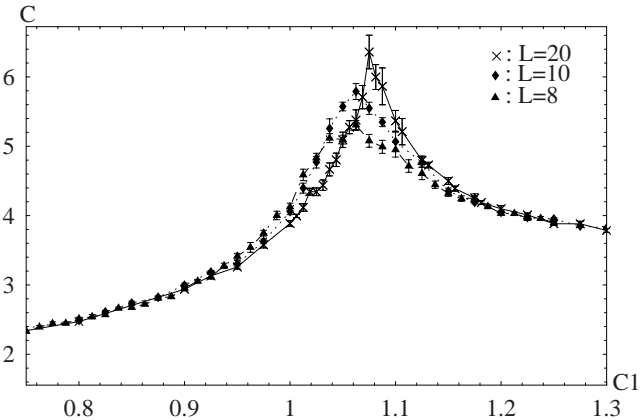}
\caption{System-size dependence of the specific heat for $c_2=2.0$.}
\label{fig.sh2}
\end{center}
\end{figure}

We first calculated thermodynamic quantities per site
like the internal energy $U$ and
the specific heat $C$,
\begin{eqnarray}
U &=& \langle S \rangle/N, \nonumber\\
C &=& \langle (S- \langle S \rangle)^2 \rangle/N, 
\end{eqnarray}
as functions of the parameters $c_1$ and $c_2$.
The specific heat $C$ has a peak as a function of $c_1$
for each fixed $c_2$. See Fig.\ref{fig.sh1} for 
$c_2 = 0$ and Fig.\ref{fig.sh2} for $c_2 = 2.0$.
These peaks develop  as the lattice 
size increases as $N=6^3,\;
10^3$, $20^3$ and $30^3$ for the $c_2=0$ case and $N=8^3,\;
10^3$ and $20^3$ for the $c_2=2.0$, respectively. 
This behavior indicates second-order phase transitions.
Fig.\ref{fig.sh1} shows that the O(3) spin model of Eq.(\ref{I0})
has a phase transition at $c_{1c}({\rm O(3)})=2.85...$.

%---------------------------------------------------
%----------------------------------------------------
%---------------------------------------------------
%----------------------------------------------------

\begin{figure}[htbp]
\begin{center}
\leavevmode
\epsfxsize=8.5cm
\epsffile{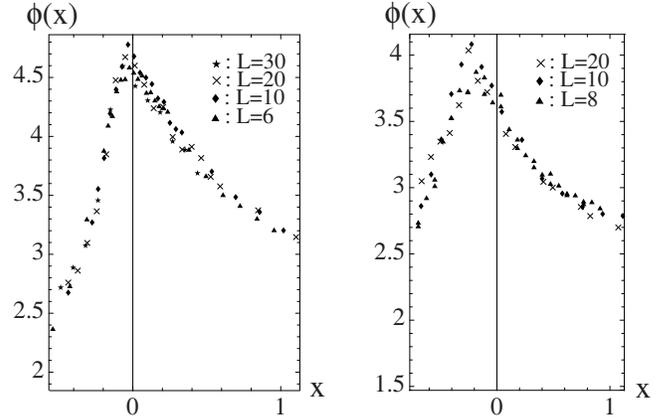}
\caption{Scaling function $\phi(x)$ obtained from the data of the
$c_2=0$ (left) and $c_2=2.0$ (right) cases.}
\label{fig.fss}
\end{center}
\end{figure}

It is interesting to see if the data of the specific heat exhibit
the finite-size scaling (FSS) behavior. 
Let us introduce a parameter $\epsilon$ as
\begin{equation}
\epsilon\equiv {c_1-c_{1\infty} \over c_{1\infty}},
\label{epsilon}
\end{equation}
where $c_{1\infty}$ is the critical coupling at the infinite volume limit.
The correlation function $\xi$ behaves as $\xi \propto \epsilon^{-\nu}$
with a critical exponent $\nu$.
It is also expected that the peak of $C$ diverges as $C_\infty \propto
\epsilon^{-\sigma}$ as $N\rightarrow \infty$ with another exponent $\sigma$.
Then for sufficiently large systems, we expect
\begin{equation}
C_N(\epsilon)=N^{{\sigma /\nu}}\phi (N^{{1/\nu}}\epsilon),
\label{FFS}
\end{equation}
where $C_N(\epsilon)$ is the specific heat in the system of size $N$ and
$\phi(x)$ is a scaling function.
We show the scaling functions $\phi$ in Fig.\ref{fig.fss}
which are determined from the data in 
Figs.\ref{fig.sh1} and \ref{fig.sh2}.
The parameters of the FSS are $c_{1\infty}=2.87,\; \nu=1.50,\;
\sigma=0.20$ for $c_2=0$ and $c_{1\infty}=1.09,\; \nu=1.17,\;
\sigma=0.17$ for $c_2=2.0$, respectively.
We think that the finite-size scaling hypothesis (FSSH) holds quite well for
the $c_2=0$ case.
For the $c_2=2.0$ case, data exhibit the FSS in 
smaller parameter region close to the peaks of $C$ than that of the
$c_2=0$ case.
We think that it is due to the existence of another length scale
besides the spin correlation length $\xi$, i.e.,
the correlation length of the gauge boson.
See the discussion in Sec.7.

\begin{figure}[htbp]
\begin{center}
\leavevmode
\epsfxsize=8.5cm
\epsffile{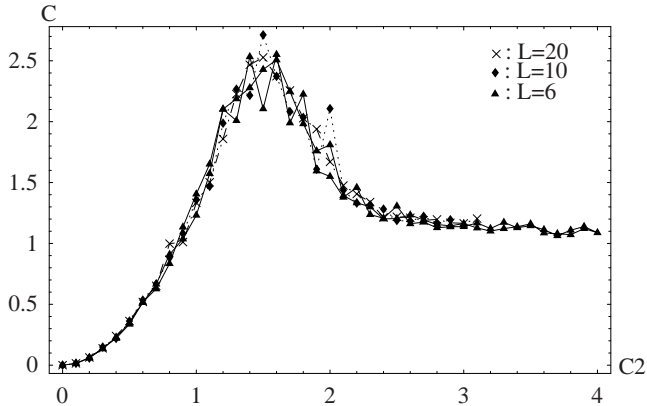}
\caption{System-size dependence of the specific heat peak 
for $c_1=0$. No developments of peaks are observed.}
\label{fig.sh3}
\end{center}
\end{figure}

Next, in Fig.\ref{fig.sh3}, we plot  $C$ at $c_1=0$
as a function of $c_2$.  It has certainly a peak at  $c_2=1.4...$, 
which, however, does not develop as the lattice size increases.  
Thus this peak does not indicates any phase transition along $c_1=0$.
This nondeveloping  peak at ($c_1=0, c_2=1.4...$)
continues to the region $c_1 > 0$.
This line of the peaks may exhibit some crossover.
The physical meaning of this line will be discussed after calculating 
instanton density in Sec.6.

%---------------------------------------------------

%----------------------------------------------------
%---------------------------------------------------
%----------------------------------------------------

\begin{figure}[htbp]
\begin{center}
\leavevmode
\epsfxsize=8cm
\epsffile{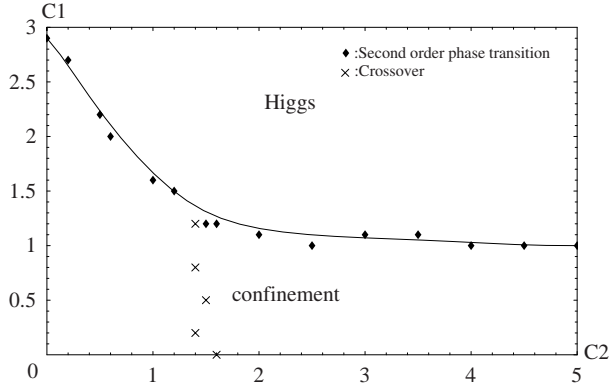}
\caption{The phase structure of CP$^1+U(1)$ model in the $c_1$-$c_2$ plane
obtained by the measurement of the specific heat.}
\label{fig.ps}
\end{center}
\end{figure}

In Fig.\ref{fig.ps} we present the phase diagram 
in the $c_2-c_1$ plane obtained via these 
specific-heat measurements. We also show the locations
of the nondeveloping peaks as crossovers.

%---------------------------------------------------
%----------------------------------------------------

%%%%%%%%%%%%%%%%%%%%%%%%%%%%%%%%%%%%%%%%%%%%%%%%%%%%%%%%%%%%%%%%%%%%%
\section{Spin excitations}

In this section we study the spin-spin correlation functions  
and the energy gap of spin excitations.
The correlator of O(3) spins, $\vec{S}_x$ of Eq.(\ref{o3spin}), 
is defined as 
\begin{equation}
C_S(t)=\langle \vec{S}_{x}\cdot \vec{S}_{x+t\mu}\rangle.
\label{spinc}
\end{equation} 
$C_S(t)$ is a function of the distance $t$ but does not depend on 
the reference point $x$ nor the direction $\mu$ due to the PBC 
on the cubic lattice.

\begin{figure}[htbp]
\begin{center}
\leavevmode
\epsfxsize=8cm
\epsffile{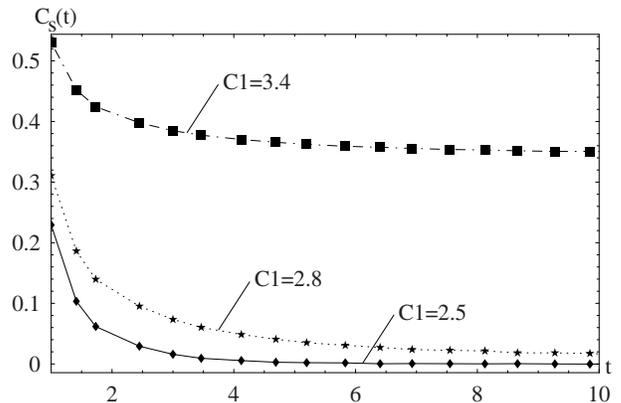}
\caption{Spin-spin correlations for $c_2=0$. 
For $c_1=2.5$, an exponential decay $A' \exp^{-\gamma' t}$ fits it better.
For $c_1=2.8$ and $3.4$, an  algebraic decay $M^2 +At^{-\alpha'}$ 
fits it better.}
\label{fig.ss1}
\end{center}
\end{figure}

\begin{figure}[htbp]
\begin{center}
\leavevmode
\epsfxsize=8cm
\epsffile{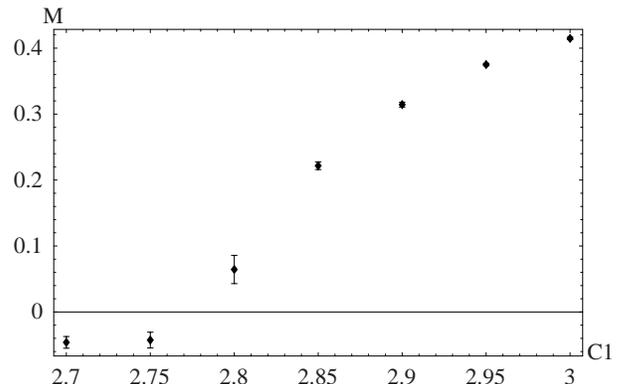}
\caption{Spontaneous magnetization for $c_2=0$. 
The result shows a typical behavior of second-order
phase transition.}
\label{fig.sm}
\end{center}
\end{figure}

Let us first consider the case $c_2=0$.
As explained before, this case
corresponds to the nearest-neighbor spin model (\ref{I0}).
In Fig.\ref{fig.ss1} we plot $C_S(t)$  for 
the lattice size $20^3$.  
$C_S(t)$ changes its behavior around $c_1=c_{1c}({\rm O(3)})=2.8...$,
the critical point determined by the location of specific heat of 
Fig.\ref{fig.sh1}.
At a higher $c_1=3.4\ [> c_{1c}({\rm O(3)})]$, 
$C_S(t)$ exhibits a nonvanishing off-diagonal long-range order 
(magnetization) $M$,
$M^2 \equiv \lim_{t\rightarrow \infty} C_S(t)$,
where $t\rightarrow \infty$ implies $t= L/2$ for a finite lattice
of size $N=L^3$. $C_S(t)$ approaches to $M^2$ algebracally,
\begin{eqnarray}
C_S(t) &=& M^2 + A t^{-(1+\eta')},\nonumber\\
 \eta' &=& 0.0\sim 0.1\ {\rm for}\ L=20. 
\end{eqnarray}
In contrast, at a lower $c_1=2.5\ [< c_{1c}({\rm O(3)})]$, 
$M=0$ and $C_S(t)$ decays exponentially,
\begin{eqnarray}
C_S(t) &=&  A' \exp(-\gamma' t).
\end{eqnarray}
Thus the phase transition observed in the previous section
corresponds to a spontaneous symmetry breaking of the SU(2) spin 
symmetry (\ref{su2spin}) of $z_x$, say $\langle z_{x} \rangle 
= (M^{1/2},0)^t$. To confirm this observation, 
we plot the magnetization
$M$ for $c_2=0$ in Fig.\ref{fig.sm}.
It exhibits
a typical behavior of second-order phase transition.

%---------------------------------------------------
%----------------------------------------------------

%---------------------------------------------------
%----------------------------------------------------

\begin{figure}[htbp]
\begin{center}
\leavevmode
\epsfxsize=8cm
\epsffile{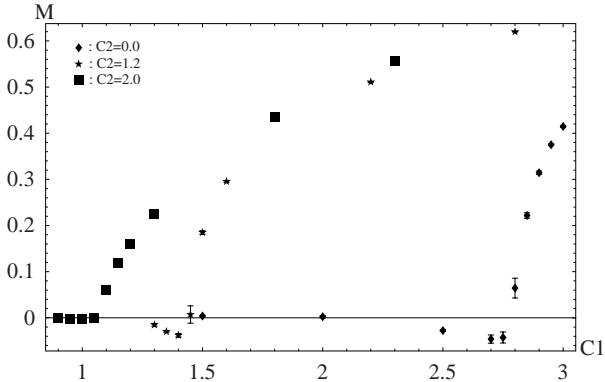}
\caption{Spontaneous magnetization for various $c_2$. 
The result shows a typical behavior of second-order
phase transition.}
\label{fig.sm2}
\end{center}
\end{figure}

In Fig.\ref{fig.sm2} we also show the magnetization $M$ 
at various values of $c_2$.
$M$ starts to develop at the critical points
observed by the specific heat measurement.
This means that $M$ is an order parameter for the
phase transition in the present model.

%---------------------------------------------------
%----------------------------------------------------

\begin{figure}[htbp]
\begin{center}
\leavevmode
\epsfxsize=8cm
\epsffile{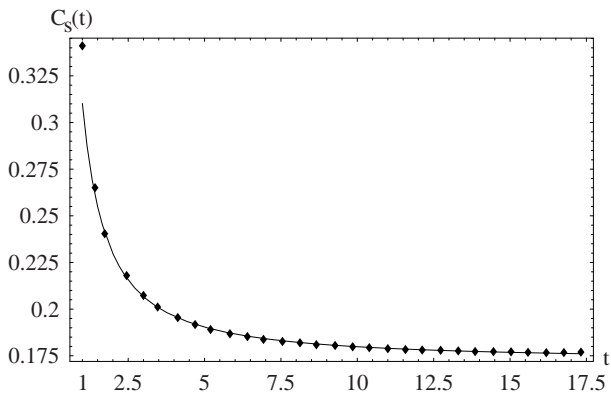}
\caption{Spin correlation function in the O(4) model
at $c_1=1.60$ and $c_2=\infty$. The system size $L=20$,
and the exponent is estimated as $\eta"=0.26$.
}
\label{fig.o3s}
\end{center}
\end{figure}

As we mentioned in Sec.2, the model reduces to the O(4) spin model 
in the limit $c_2 \rightarrow \infty$.
We studied the phase structure of this O(4) model, 
and found that there 
exists a second-order order-disorder phase transition as 
in the pure CP$^1$ model.
The critical coupling is estimated as 
$c_{1c}({\rm O(4)})=1.0...$ which is
fairly close to the critical value of $c_1$ for sufficiently 
large $c_2$.
We also investigate the spin correlation function in the 
ordered phase.
The data can be fitted well in the form (see Fig.\ref{fig.o3s}),
\begin{eqnarray}
C_S(t)&=&M^2 +b t^{-(1+\eta")},\nonumber\\
\eta"&=& 0.1\sim 0.3\ {\rm for}\ L=20.
\label{CS2}
\end{eqnarray}
This result indicates that smooth transfer in the behavior
of $C_S(t)$ occurs 
from the O(3) spin model at $c_2\simeq 0$ to
the O(4) spin model at $c_2\simeq \infty$.

%---------------------------------------------------
%----------------------------------------------------

Next let us study the mass (energy gap) $M_S$ of  spin excitations. 
To measure $M_S$ with good precision, 
we introduce a Fourier transform of the spin variables $\vec{S}_x$ 
in the 2D plane  $(x_1,x_2)$%\cite{fourier},
\begin{equation}
\tilde{\vec{S}}(x_3;p_1,p_2)=\sum_{x_1,x_2}e^{ip_1x_1+ip_2x_2}
\vec{S}_x.
\label{FS}
\end{equation}
In the continuum limit, one can readily find that the correlator of 
$\tilde{\vec{S}}(x_3;p_1,p_2)$ behaves as
\begin{eqnarray}
\langle \tilde{\vec{S}}(x_3;p_1,p_2)\cdot \tilde{\vec{S}}
(0;-p_1,-p_2)\rangle
&=& \int dp_3 {e^{ip_3x_3}\over \vec{p}^2+M_S^2}  \nonumber \\
&\propto& e^{-\sqrt{p_1^2+p_2^2+M_S^2}x_3},
\label{CFS}
\end{eqnarray}
where $\vec{p}^2=\sum_{i=1,2,3}p^2_i$ and $M_S$ is the
energy gap of the lowest spin excitations. 
Below we fit the measured
correlation function on the lattice 
by using this form to calculate $M_S$. Let us define 
\begin{equation}
D_S(t)={1\over L^3}\sum_{x_3}\langle 
\tilde{\vec{S}}(x_3+t;p_1,p_2)\cdot 
\tilde{\vec{S}}(x_3;-p_1,-p_2)\rangle.
\label{DS}
\end{equation}
The PBC restricts 
the momentum as $p_i = 2\pi n_i/L,
\ n_i= 0,1,2,\cdots, L-1$. In the simulation we choose nonvanishing
smallest values, 
 $p_1=p_2=2\pi/L$ in order to fit the result in  an
 exponentially-decaying form even for $M_S=0$.
We note that, from the above definition of $M_S$, it is possible 
that $M_S^2$ takes a negative value as long as 
the exponent $p_1^2+p_2^2+M_S^2 < 2(2\pi/L)^2$. 
We shall comment on this point later on.  

\begin{figure}[htbp]
\begin{center}
\leavevmode
\epsfxsize=8cm
\epsffile{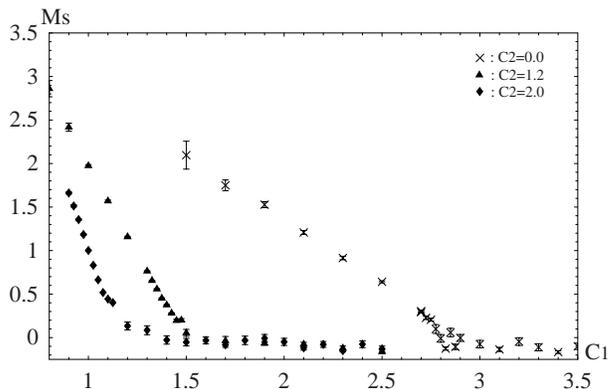}
\caption{Energy gap $M_S$ of O(3) spin excitations 
for $c_2=0.0,1.2$ and $2.0$. 
The errors become larger for smaller $c_1$ because of
more rapid fall off of $D_S(t)$.}
\label{fig.ss2}
\end{center}
\end{figure}

In Fig.\ref{fig.ss2}, we plot $M_S$ for $N=20^3$.
For a fixed $c_2$, $M_S$ is a smooth function of $c_1$ and 
vanishes in the spin ordered phase $c_1 > c_{1c}$
as predicted by the Nambu-Goldstone theorem.
We checked the result of Fig.\ref{fig.ss2} by repeating
calculation with the choice $(p_1, p_2)=(2\pi/L,0)$.
We shall use the same techniques for calculating 
the gauge-boson mass in the following section.

%---------------------------------------------------
%----------------------------------------------------

%%%%%%%%%%%%%%%%%%%%%%%%%%%%%%%%%%%%%%%%%%%%%%%%%%%%%%%%%%%%%%%%%%%%%%%%

\section{Instantons}
\setcounter{equation}{0}

\subsection{Definition of instanton density}

In this section, we calculate the instanton 
density $Q_x$ at $x$ in various set of ($c_2, c_1$).
Measurement of $Q_x$ provides us with important information
about fluctuations of the gauge field, though it is rather difficult
to determine a location of phase transition by itself.

We follow the definition of integer instanton charge by DeGrand and 
Toussaint\cite{instanton}.
First, let us consider the magnetic flux $\Theta_{x,\mu\nu}$
penetrating plaquette $(x,x+\mu,x+\mu+\nu,x+\nu$),
\begin{eqnarray}
&& \Theta_{x,\mu\nu}\equiv \theta_{x\mu}+\theta_{x+\mu,\nu}
-\theta_{x+\nu,\mu}-\theta_{x\nu}, \nonumber \\
&& \hspace{2cm} (-4\pi<\Theta_{x,\mu\nu}<4\pi).
\label{Theta}
\end{eqnarray}
We decompose $\Theta_{x,\mu\nu}$ into its 
{\it integer} part $2\pi n_{x,\mu\nu}$ ($n_{x,\mu\nu}$ is an integer)
and the remaining part  $\tilde{\Theta}_{x,\mu\nu} \equiv$
 $\Theta_{x,\mu\nu}\; (\mbox{mod} \;2\pi$) uniquely,
\begin{equation}
\Theta_{x,\mu\nu}=2\pi n_{x,\mu\nu}+\tilde{\Theta}_{x,\mu\nu}, \;\;
(-\pi<\tilde{\Theta}_{x,\mu\nu}<\pi).
\end{equation}
Physically, $n_{x,\mu\nu}$ describes the 
Dirac string whereas $\tilde{\Theta}_{x,\mu\nu}$ describes the
fluctuations around it.
The quantized instanton charge $Q_x$ at the cube 
around the site $\tilde{x} =x+{\hat{1} \over 2}
+{\hat{2} \over 2}+{\hat{3} \over 2}$ of the dual lattice 
is defined as 
\begin{eqnarray}
Q_x&=&
-{1\over 2}\sum_{\mu,\nu,\rho}\epsilon_{\mu\nu\rho}
(n_{x+\mu,\nu\rho}-n_{x,\nu\rho})\nonumber\\
&=&{1\over 4\pi}\sum_{\mu,\nu,\rho}\epsilon_{\mu\nu\rho}
(\tilde{\Theta}_{x+\mu,\nu\rho}-\tilde{\Theta}_{x,\nu\rho}),
\label{instden}
\end{eqnarray}
where $\epsilon_{\mu\nu\rho}$ is the complete antisymmetric tensor.
$Q_x$ measures the total flux emanating
from the monopole(instanton) sitting at $\tilde{x}$. 
Roughly speaking, $Q_x$ measures the strength of nonperturbative
gauge configurations. Polyakov showed that a condensation
of these instantons drives the system at $c_1=0$ into a confinement phase,
which is characterized by strong and large fluctuations of gauge field
$\theta_{x\mu}$ at long distances. 

%%%%%%%%%%%%%%%%%%%%%%%%%%%%%%%%%%%%%%%%%%%%%%%%%%%%%%%%%%%%%%%%%%%%
\subsection{Results}

\begin{figure}[htbp]
\begin{center}
\leavevmode
\epsfxsize=8cm
\epsffile{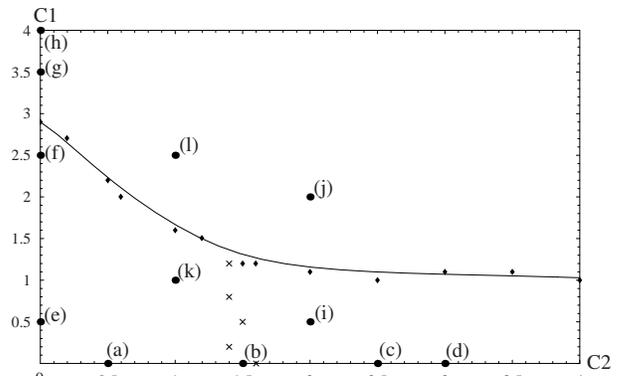}
\caption{Instanton distribution is measured at various places
(a) $\sim$ (l) in the phase diagram in the $c_2-c_1$ plane.}
\label{fig.ins1}
\end{center}
\end{figure}

\begin{figure}[htbp]
\begin{center}
\leavevmode
\epsfxsize=12cm
\epsffile{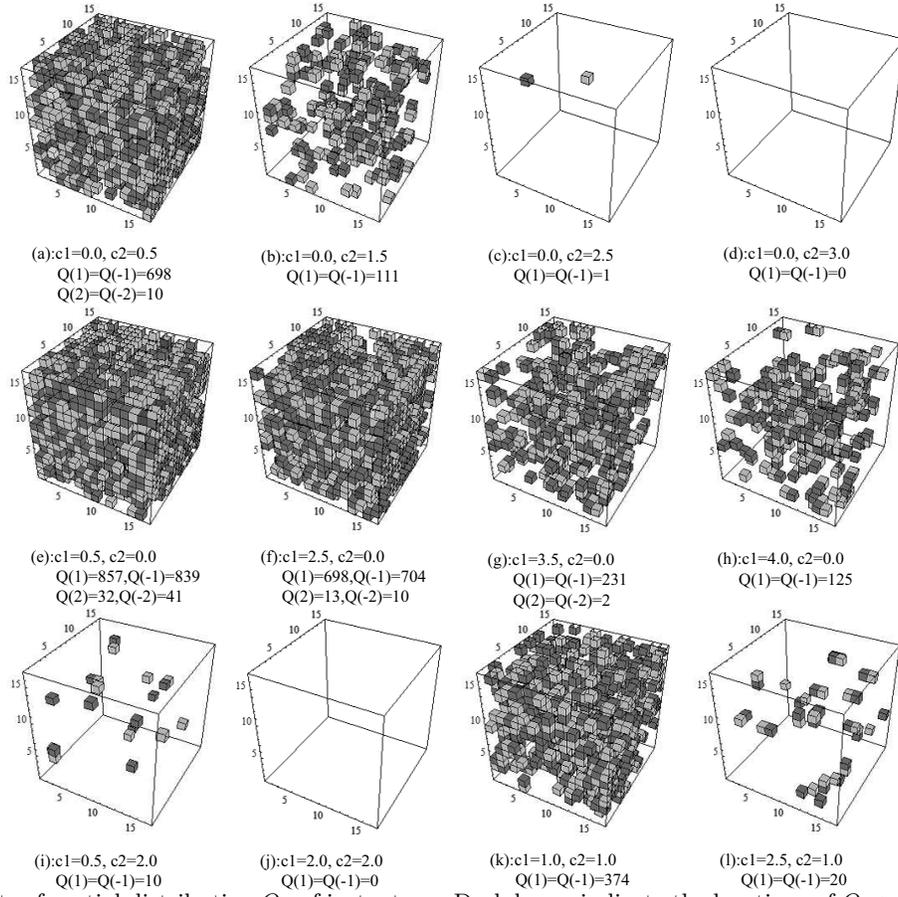}
\caption{Snapshots of spatial distribution $Q_x$ of instantons.
Dark boxes indicate the locations of $Q_x >0$  whereas
light boxes indicate $Q_x < 0$. $Q(n) =  \sum_x
\delta_{n, |Q_x|}  $ is the density of instantons 
with the charge $n$.}
\label{fig.ins2}
\end{center}
\end{figure}

\begin{figure}[htbp]
\begin{center}
\leavevmode
\epsfxsize=8cm
\epsffile{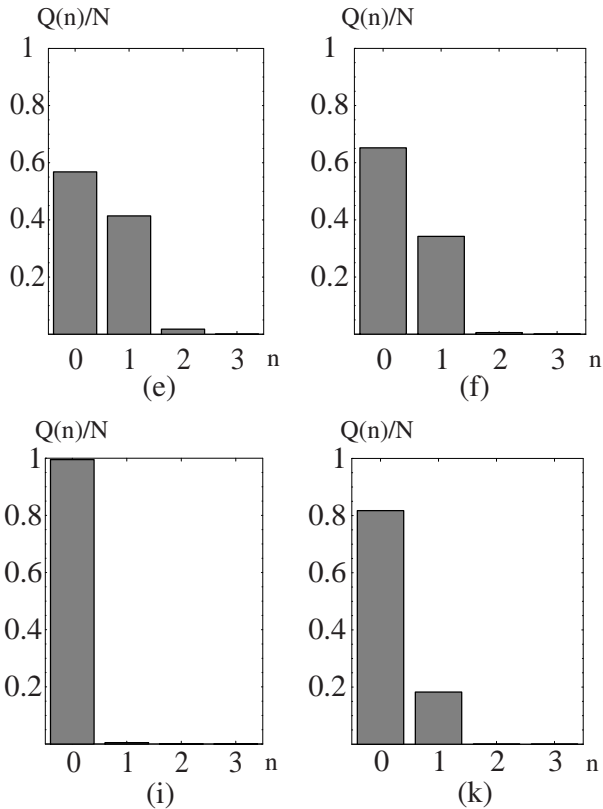}
\caption{Distribution of the instanton charge density $Q(n)$ at the 
locations $(e)-(k)$ in the phase diagram. $Q(n\geq 3)=0.$}
\label{fig.qbars}
\end{center}
\end{figure}

In this subsection, we shall show ``snapshots" of $Q_x$,
typical configurations of $Q_x$ taken
 in the process of  MC updates after sufficient 
 thermalization.
In Fig.\ref{fig.ins1}, we first show the locations of
pairs of $(c_2,c_1)$ in the phase diagram
at which snapshots are taken.
In Fig.\ref{fig.ins2}, we present the snapshots at these locations
for the lattice $N=16^3$. 
In Fig.\ref{fig.qbars},
we show $Q(n)$, the density of instantons 
with the charge $n (\geq 0)$, $Q(n) =  \sum_x\delta_{n, |Q_x|}/N $.
The length of each hexahedron shows the magnitude of 
instanton charges there.

Let us first see the behavior of instantons 
in the pure $3$D U(1) gauge system with $c_1=0$ [(a),(b),(c) and 
(d) in Fig.\ref{fig.ins1}].
The result indicates that the instanton density decreases very 
rapidly from $c_2 \sim 1.5$ as $c_2$ increases.
Therefore, we can identify the crossover found by  
the specific heat $C$ in Sec.4
as a crossover from the region of dense instantons
to that of dilute instantons.
In fact this was first observed in Ref.\cite{cross}.

%---------------------------------------------------
%----------------------------------------------------

\onecolumn

%---------------------------------------------------
%----------------------------------------------------

\twocolumn

%---------------------------------------------------
%----------------------------------------------------

\begin{figure}[htbp]
\begin{center}
\leavevmode
\epsfxsize=8cm
\epsffile{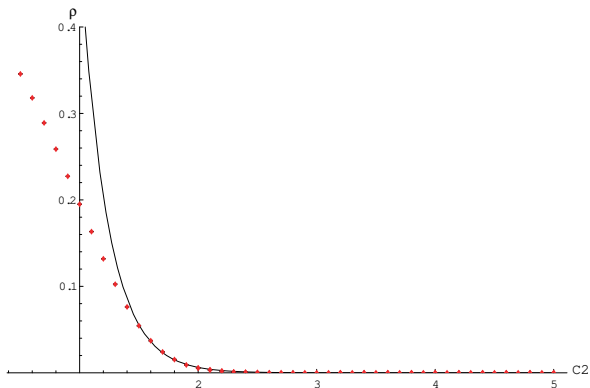}
\caption{Average of instanton density $\rho = \langle \sum_x|Q_x|
\rangle/N$ for $c_1=0$ as a function of $c_2$.
The solid curve  is a fit in the form of dilute-gas
approximation $\propto \exp(-\ell c_2)$ with $\ell \simeq 4.35$.}
\label{fig.ins_fit}
\end{center}
\end{figure}

More systematically, we calculate the average density,
$\rho \equiv \langle \sum_x |Q_x| \rangle/N$ for $c_1=0$ as a function
$c_2$. See Fig.\ref{fig.ins_fit}.
We fit the result with $\exp(-\ell c_2)$ as expected by the dilute-gas 
approximation\cite{polyakov}, where $\ell\ (\simeq 4.35)$, 
whereas its theoretical value is calculated as 
5.06 \cite{polyakov}. 
The fitting is quite satisfactory for $c_2>1.5$ as expected.

%---------------------------------------------------
%----------------------------------------------------

Let us turn to the pure CP$^1$ model at $c_2=0$.
In Fig.\ref{fig.ins2}, the snapshots ((e) $\sim$ (h)) represent
this case.
The $c_1$-term of the action shows that a spin order, i.e., 
a condensation of $z_x$, suppresses 
the fluctuations of $U_{x\mu}$.
Due to this correlation $\bar{z}_{x+\mu}z_x \leftrightarrow U_{x\mu}$,
instanton configurations in (2+1)-dimensions 
correspond to creations and/or destructions of so-called skyrmions 
configurations of O(3) spins in the 2D plane.
Therefore instantons  are suppressed in the spin-ordered state. 
The results in Fig.\ref{fig.ins2} are consistent 
with the above observation 
as the instanton density decreases as $c_1$ increases.
However, we note that even in the spin-ordered phase at
$c_1=3.5$ there are a number of surviving instantons, 
most of which make instanton-anti-instanton dipoles 
in the nearest-neighbor pairs.

Finally, let us take a look at the snapshots 
obtained in the region close to
the critical line; the snapshots at (i) $\sim$ (l) 
in Fig.\ref{fig.ins2}.
We see that the instanton density 
increases as $c_1$ and/or $c_2$ decrease.
Even in the spin-ordered phase, the instanton density is finite for
small $c_2$. 
We do {\em not} think that this result contradicts a natural 
belief that
the Higgs phase is realized in the spin-ordered phase.
For the confinement phase to appear, finite instanton-density
is not sufficient, but dissociations of 
instanton-anti-instanton dipoles are necessary.

The observation of instantons in the present section is 
quite helpful for understanding the behavior of gauge-boson mass, 
which is calculated in the following section.

%%%%%%%%%%%%%%%%%%%%%%%%%%%%%%%%%%%%%%%%%%%%%%%%%%%%%%%%%%%%%%%%%%%%%%%%
\section{Mass of gauge bosons}
\setcounter{equation}{0}

In this section we study the gauge-invariant mass $M_G$ of
gauge bosons. 
The present model contains two independent dynamical
variables, i.e., the CP$^1$ field and the gauge field,
and therefore there exist two independent length scales.
In Sec.5, we studied the correlation length of the CP$^1$
spin sector.
In this section, we investigate that of the gauge sector.

We explain the definition of $M_G$ in Sec.7.1, 
and show the numerical result of $M_G$ in Sec.7.2.
In Sec.7.3 we address the problem of whether
the system is in the confinement phase or in Coulomb phase 
just on the critical
points $c_{1c}$. In Sec.7.4, we study the case of noncompact
gauge field to support our argument for the problem in Sec.7.3.

\subsection{Definition of gauge-boson mass $M_G$}

In this section, we study the gauge-invariant mass of gauge bosons,
$M_G$, by using the same techniques %\cite{fourier} 
 that we used to measure the spin gap $M_S$ in the previous section.
To define $M_G$ we first introduce a gauge-invariant operator $O(x)$,
\begin{eqnarray}
O(x)&=&\sum_{\mu,\nu=1,2}
\epsilon_{\mu\nu}\mbox{Im} \bar{U}_{x\nu}\bar{U}_{x+\nu,\mu}
U_{x+\mu,\nu}U_{x\mu}  \nonumber  \\
&=&\sum_{\mu,\nu}
\epsilon_{\mu\nu}\sin (-\theta_{x\nu}-\theta_{x+\nu,\mu}
+\theta_{x+\mu,\nu}+\theta_{x\mu}).
\label{O}
\end{eqnarray}
Then we intorduce a Fourier transform as before,
\begin{equation}
\tilde{O}(x_3)=\sum_{x_1,x_2}O(x)e^{ip_1x_1+ip_2x_2}.
\label{tO}
\end{equation}
to define the gauge correlation function as
\begin{equation}
D_G(t)={1\over L^3}\sum_{x_3}\Big\langle 
\tilde{O}(x_3)\bar{\tilde{O}}(x_3+t)\Big\rangle.
\label{CG}
\end{equation}
In the continuum, $D_G(t)$ is expected to behave as
\begin{eqnarray}
D_G(t) &=& \int dp_3 {e^{ip_3t}\over \vec{p}^2+M_G^2} \nonumber \\
&\propto& e^{-\sqrt{p_1^2+p_2^2+M_G^2}t}.
\label{CG2}
\end{eqnarray}
We determine $M_G$ by fitting the data in this exponential 
form (\ref{CG2}). For practical calculations, 
we set $p_1=p_2=2\pi/L$ as before.
Here we note that we are assuming the ordinary form of the 
gauge-boson propagator for a gauge system coupled with 
{\it massive} matter fields.
{\em On the critical line}, the gauge-boson propagator may be 
modified due to the appearance of massless field $z_{xa}$.
See later discussion on this point. 

\begin{figure}[htbp]
\begin{center}
\leavevmode
\epsfxsize=8cm
\epsffile{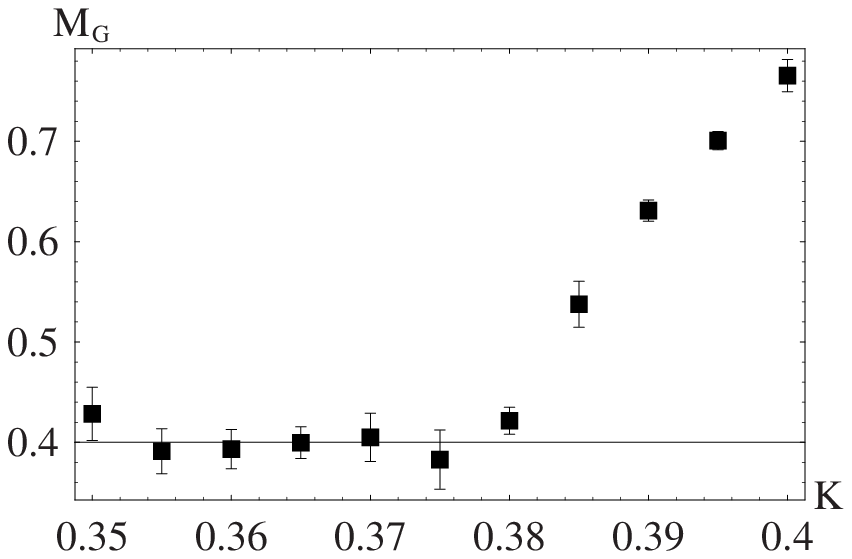}
\caption{Gauge-boson mass of Eq.(\ref{GH1}) 
for $\beta=2$ and $\lambda=0.020$,
the same parameters as in Ref.\cite{GH1}.}
\label{fig.hm1}
\end{center}
\end{figure}

In order to verify that the above methods give correct results, 
we first apply them to the 3D  U(1) lattice gauge-Higgs
models for which the gauge-boson mass has been already calculated
by using other methods.
In Ref.\cite{GH1} the action $S$ of the model is parameterized 
as follows;
\begin{eqnarray}
S_{Ch}&=&-{\beta \over 2} \sum_P\prod U-
{K \over 2}\sum_{x,\mu}\left(\bar{\Phi}_{x+\mu}U_{x\mu}
\Phi_x +\mbox{H.c.}\right)\nonumber \\
&&+\sum_x\left[|\Phi_x|^2+\lambda(|\Phi_x|^2-1)^2\right],
\label{GH1}
\end{eqnarray}
where $\Phi_x$ is the complex Higgs field.
By changing the value of $K$, the system undergoes a 
Higgs-confinement phase transition.
A gauge-boson mass was calculated by studying the {\em genuine 
gauge-boson propagator} with a specific gauge-fixing condition.
We calculated the gauge boson mass for the same parameter region 
as in Ref.\cite{GH1} but by using $D_G(t)$ of Eq.(\ref{CG}).
In this region, the gauge coupling constant has a moderate value
$\beta=2$ and $\lambda=0.020$.
In Fig.\ref{fig.hm1} we show our results of the gauge-boson mass, 
which are in good agreement with those of Ref.\cite{GH1}.
We should notice that 
the gauge boson mass is always nonvanishing even at the critical 
point, though one might expect the behavior $M^2_G
\sim |K-K_c|$ where $K_c$ is the value at the critical point.
We shall comment on this point after showing the results of the 
present CP$^1$+U(1) model.

\begin{figure}[htbp]
\begin{center}
\leavevmode
\epsfxsize=8cm
\epsffile{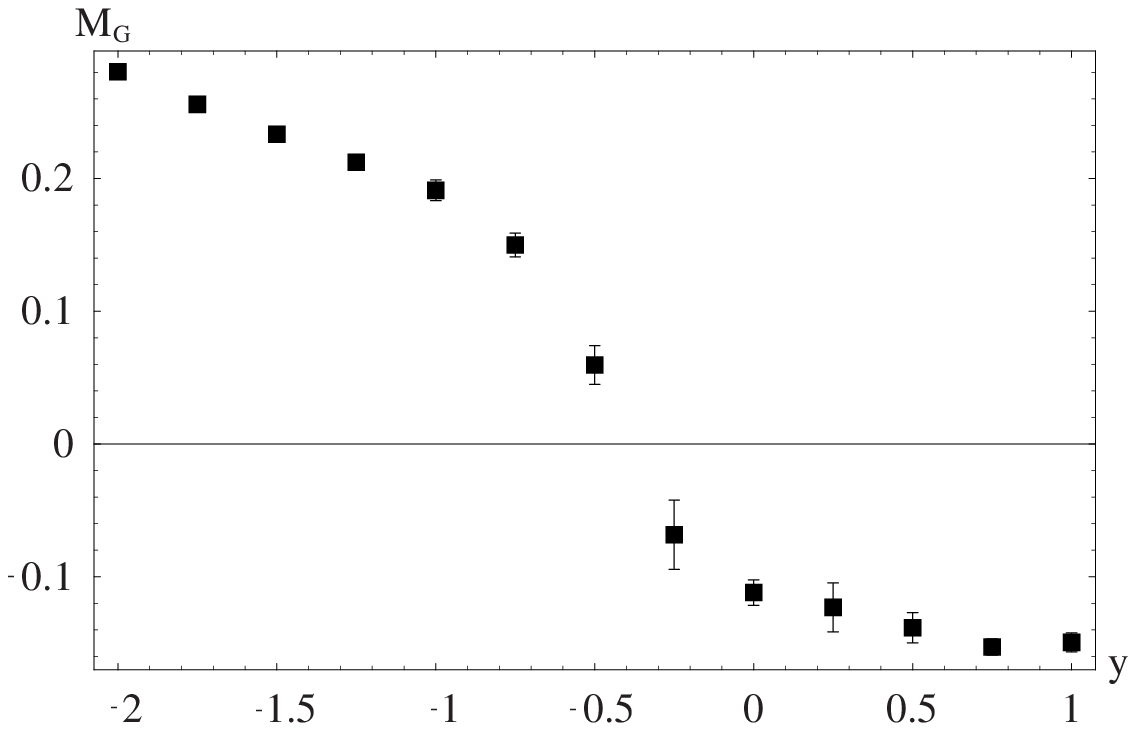}
\caption{Gauge-boson mass of Eq.(\ref{GH2}) 
for $x=2,\beta=4$ and $K=5$, 
the same parameters as in Ref.\cite{GH2}.
Here and in the forthcoming figures of $M_G$, $M_G$ denotes the product
sgn$(M_G^2)\cdot \sqrt{|M_G^2|} $ with $M_G^2$ defined by
Eq.(\ref{CG2}) in order to distinguish the case of negative $M_G^2$
clearly.
}
\label{fig.hm2}
\end{center}
\end{figure}

On the other hand, in Ref.\cite{GH2}, it was reported that
the gauge-boson mass vanishes in a certain parameter region. 
In Ref.\cite{GH2}, the Higgs part of the action is parameterized 
in a complicated way as
\begin{eqnarray}
S_{Ka}&=&-{\beta \over 2}\sum_P\prod U-{K \over 2}\sum_{x,\mu}
\left(\bar{\Phi}_{x+\mu}U_{x\mu}\Phi_x
+\mbox{H.c.}\right) \nonumber  \\
&&+{K \over 2}\sum_x |\Phi_x|^2\Big[6+{y\over K^2}-
{3.17(1+2x) \over 2\pi K} \nonumber \\
&&-{(-4+8x-8x^2)(\log 6\beta+0.09)+25.5+4.6x \over 
16\pi^2 \beta^2} \Big] 
\nonumber \\
&&+{xK^2 \over 4\beta}\sum_x |\Phi_x|^4.
\label{GH2}
\end{eqnarray}
We  also calculated the gauge-boson mass of Eq.(\ref{GH2}).
In Fig.\ref{fig.hm2} we present our result of $M_G$ for
$x=2,\beta=4$ and $K=5$, the same parameters used in Ref.\cite{GH2}.
The result agrees with that of Ref.\cite{GH2}.
For positive $y$, $M_G$ reduces to around zero.
This phenomenon may be understood naturally as follows; 
For large $y$
the mass term of $\Phi_x$ becomes large 
and $\Phi_x$ fluctuates weakly around zero. Then one can treat
the hopping $K$ term as  
a small perturbation. The main term is the first  
$\beta$-term whose coefficient is large.
Thus the fluctuations of gauge bosons are strongly suppressed,
so one can expand as $U_{x\mu}\simeq  1+ iA_{x\mu}$. 
The resulting action $\propto \Theta_{x,\mu\nu}^2$ describes
just free massless gauge bosons.
See also the discussion in the following subsection.

%---------------------------------------------------
%----------------------------------------------------
%---------------------------------------------------
%----------------------------------------------------

%%%%%%%%%%%%%%%%%%%%%%%%%%%%%%%%%%%%%%%%%%%%%%%%%%%%%%%%%%%%%%%%

\subsection{The case of compact U(1) gauge field}

We measured the gauge-boson mass for the   $16^3$ lattice.
So the distance $t$ in $D_G(t)$ of (\ref{CG2}) 
takes $t=1, \cdots, 8$ due to 
the PBC.
The value of $M_G$ for each point of the $c_2-c_1$ plane
is calculaed  as the average  of  10 samples, each sample
of which is  obtained after 
$2\times 10^5 \sim 4 \times 10^5$ sweeps.
The error of $M_G$ is evaluated  by the standard deviation of 
these 10 samples.

There are two mechanisms to generate nonvanishing $M_G$:
(i) Condensation of instantons and (ii) Anderson-Higgs mechanism.
We shall see them first by studying  
$M_G(c_2)$ as  a function of $c_2$ for 
a fixed $c_1$, and then by studying $M_G(c_1)$ for a fixed $c_2$.

\begin{figure}[htbp]
\begin{center}
\leavevmode
\epsfxsize=8cm
\epsffile{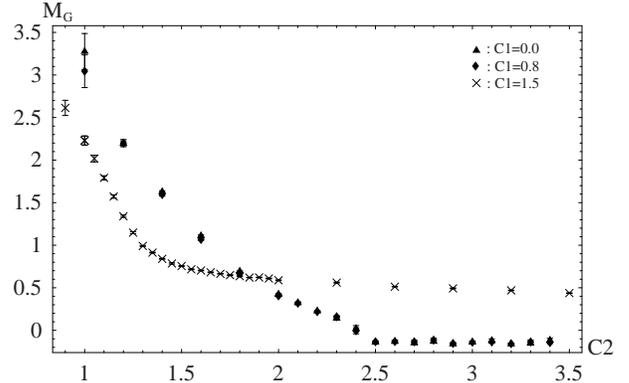}
\caption{Gauge-boson  mass $M_G$ as a function of $c_2$ for 
$c_1=0.0,0.8,1.5$. The data of $c_1=0.0$ and $0.8$ almost overlap.}
\label{fig.pm1}
\end{center}
\end{figure}

In Fig.\ref{fig.pm1} we present $M_G(c_2)$ for $c_1=0,0.8$
and $1.5$.
Let us first see the pure gauge case $c_1=0$ in some detail. 
As $c_2$ increases from zero,  $M_G$ decreases 
and becomes almost zero at $c_2 \simeq 2.5$, the value at which
the instanton density almost vanishes.
Thus we find a positive correlation between the gauge-boson mass 
and 
the instanton density. Actually, in the dilute-gas approximation,
$M_G$ is estimated as $M_G \propto \rho \propto 
\exp(-\ell c_2 )$\cite{polyakov}. 
In fact, for $\rho \simeq 0$,
one may forget the nonperturbative (instanton) 
effect and expands as $U_{x\mu}
\simeq 1+i\theta_{x\mu}$ in the action. The action of
small fluctuations $\theta_{x\mu}$ becomes
$S \simeq \sum_{x\mu\nu}\Theta_{x\mu\nu}^2$, which
describe a free massless gauge boson as explained before 
in the case of Higgs model\cite{GH2}.
On the other hand, for the system of dense instantons, 
the gauge field $\theta_{x\mu}$
has short-range, i.e., massive correlations because of their wild
fluctuations.
In short, as the $c_2$ term controls the density of instantons, 
i.e., wild fluctuations of $\theta_{x\mu}$, $M_G(c_2)$ 
decreases as $c_2$ increases.
Fig.\ref{fig.pm1} shows that this tendency survives 
for $c_1= 0.8$ and $1.5$. This is expected from Fig.\ref{fig.ins2} 
of instanton density.
We remark here that 
the vanishing gauge-boson mass in the numerical calculations 
in {\em finite systems} does {\em not} guarantee
that the system is in the {\em deconfinement} phase.
The measurement of the specific heat indicates that the system is 
in the confinement phase even for large $c_2$ if $c_1$ is below 
$c_{1c}(c_2)$.
We think that the above calculation $M_G\sim 0$ in the confinement
phase is a finite-size effect and if we calculate $M_G$ in very 
large systems, we obtain $M_G>0$ in the confinement phase.

\begin{figure}[htbp]
\begin{center}
\leavevmode
\epsfxsize=8cm
\epsffile{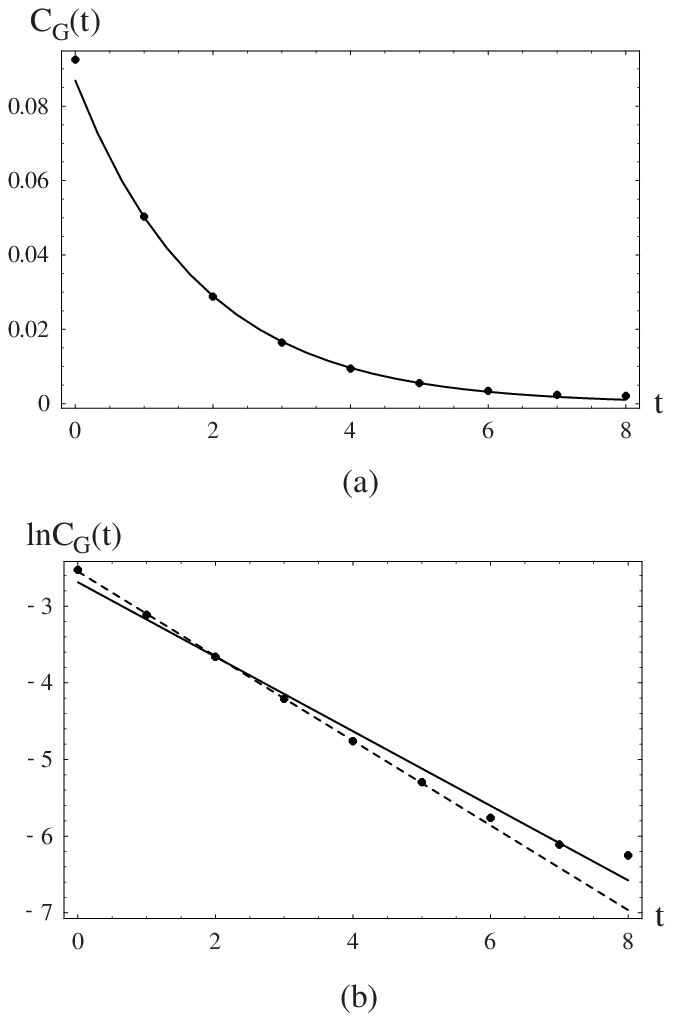}
\caption{Correlation function $D_G(t)$ of gauge boson  vs $t$
at ($c_1=0.0, c_2=3.0$) for $L=16$. (a) the exponential curve
fitting the data with $t=0$ dropped.  
(b) Log plot $\ln D_G(t)$. The solid straight line fits all
the data, while the dashed straight line fits the data
with  $t=6, 7, 8$ dropped.}
\label{fig.CG}
\end{center}
\end{figure}

In order to verify the above expectation,
let us take a look at Fig.\ref{fig.pm1} in more detail.
One sees that $M^2_G=-(0.1)^2\sim-(0.15)^2$ for $c_2>2.5$ 
with $c_1=0$ and $c_1=0.8$.
As mentioned below Eq.(\ref{DS}), these negative values of $M_G^2$
are possible in the present definition of $M_G^2$.
We think that this unphysical result stems from the finite 
size of the system as mentioned above.
Similar results are reported in Ref.\cite{EJJLN} for the 
$D=4$ compact scalar QED.
In order to see the finite-size effect on the gauge boson mass,
we plot data of $D_G(t)$ in $t-\ln D_G(t)$ plane 
in Fig.\ref{fig.CG}.
From this plot, it is obvious that the PBC 
affects the behavior of $D_G(t)$ for $t\sim L/2$ where $L$
is the system size.
Then one may discard the data of $t\sim L/2$ and re-estimate 
the gauge-boson mass to obtain $M^2_G=-(0.08)^2$.
To get a definite conclusion for the negative gauge boson mass,
calculation in larger systems is required.
In this direction, we calculated $M_G$ in a larger system  $24^3$
and obtained a reduced value $M_G^2 \sim -(0.03)^2$.
Therefore we expect that we obtain $M_G^2 \ge 0$ by calculation 
in very large
systems, and in particular $M_G^2 > 0$
in the confinement phase as it is generally expected.

%---------------------------------------------------
%----------------------------------------------------
%---------------------------------------------------
%----------------------------------------------------

On the other hand, the data for $c_1=1.5$ is interesting because
the horizontal line with $c_1=1.5$ in the $c_2-c_1$ plane
intersects the critical line
as $c_2$ is increased.
The intersection point is estimated as $(c_1=1.5,c_2=1.2)$ from the
phase diagram Fig.\ref{fig.ps}.
As seen in Fig.\ref{fig.pm1}, 
the gauge boson mass decreases very rapidly until 
$c_2\sim 1.3$ and becomes almost constant for $c_2 >1.4$.
This rapid decrease of $M_G$ in the spin-disordered phase comes
from the decrease of the instanton density as we measured in the
previous section.
The finite gauge-boson mass in the spin-ordered phase can 
be interpreted as due to the {\em Anderson-Higgs mechanism} 
because $M_G$ stays almost constant for varying $c_2$.
We shall explain this mechanism in more detail for $M_G(c_1)$ below.

\begin{figure}[htbp]
\begin{center}
\leavevmode
\epsfxsize=8cm
\epsffile{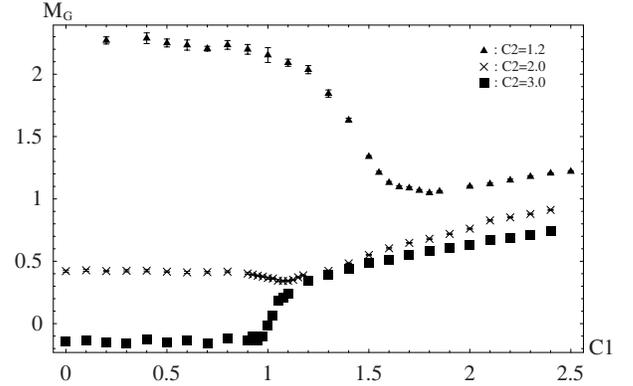}
\caption{Gauge boson mass vs  $c_1$ for 
fixed $c_2$'s.}
\label{fig.pm2}
\end{center}
\end{figure}

\begin{figure}[htbp]
\begin{center}
\leavevmode
\epsfxsize=8cm
\epsffile{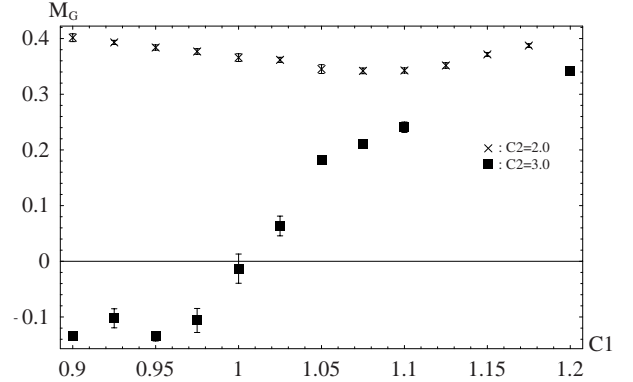}
\caption{Magnified figure of Fig.\ref{fig.pm2} near the critical
line.}
\label{fig.pm3}
\end{center}
\end{figure}

\begin{figure}[htbp]
\begin{center}
\leavevmode
\epsfxsize=8cm
\epsffile{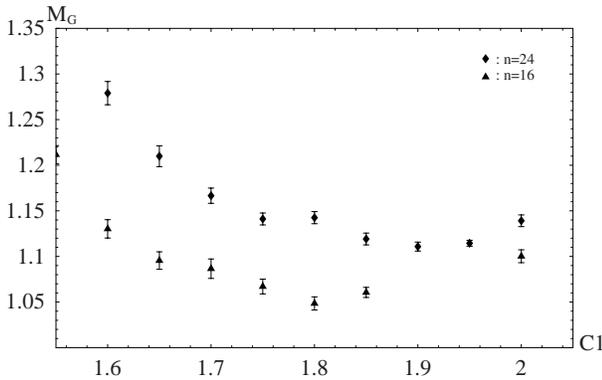}
\caption{Size dependence of the gauge-boson mass with $c_2=1.2$.}
\label{fig.pm4}
\end{center}
\end{figure}

\begin{figure}[htbp]
\begin{center}
\leavevmode
\epsfxsize=8cm
\epsffile{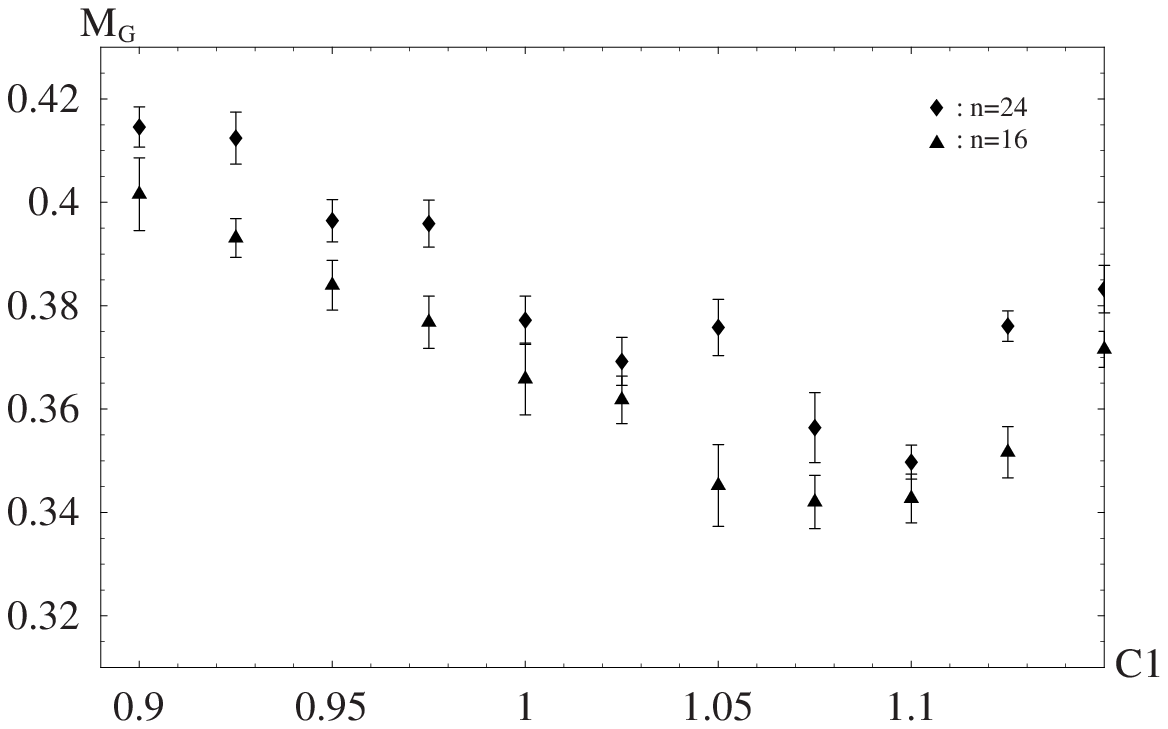}
\caption{Size dependence of the gauge-boson mass with $c_2=2.0$.}
\label{fig.pm5}
\end{center}
\end{figure}

\begin{figure}[htbp]
\begin{center}
\leavevmode
\epsfxsize=8cm
\epsffile{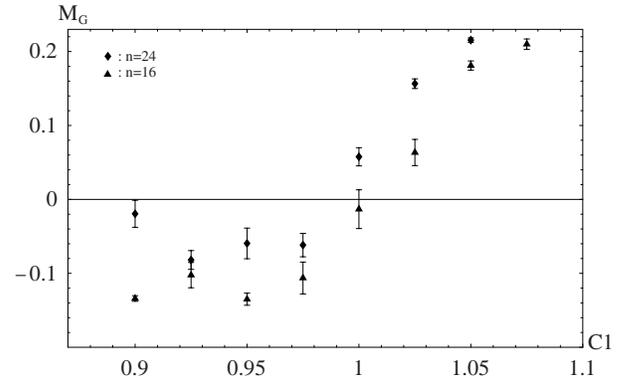}
\caption{Size dependence of the gauge-boson mass with $c_2=3.0$.}
\label{fig.pm6}
\end{center}
\end{figure}

Next, let us study $M_G(c_1)$ for a fixed $c_2$. 
In Figs.\ref{fig.pm2}-\ref{fig.pm6}, we 
present $M_G(c_1)$. 
For all values of $c_2$, as $c_1$ increases from zero,
$M_G(c_1)$ first decreases in the spin-disordered (confinement) 
phase, $0 < c_1 < c_{1c}(c_2)$, to reach the minimum at the critical point 
$c_1=c_{1c}(c_2)$.  Then it increases in the spin-ordered (Higgs) phase
$c_{1c} (c_2)< c_1$.
The first decrease is due to the suppression of instantons 
at larger $c_1$ as we observed in the previous section.
The situation  parallels  the pure gauge system $c_1=0$
if the axes $c_1$ and $c_2$ are interchanged.  
 
The successive increase  in the ordered phase is due to the 
Anderson-Higgs mechanism.
In the ordered phase,
 $\langle z_x \rangle$ starts to develop at $c_1=c_{1c}$
 as Fig.\ref{fig.sm} shows.  
Then the $c_1$ term of the action supplies the following mass term; 
\begin{eqnarray}
S_M &=& \frac{1}{2}M_{G}^2 \sum_{x\mu}\theta_{x\mu}^2, \ \ \ 
M_{G}^2 = c_1 \langle z_x \rangle^2,
\label{mftmass}
\end{eqnarray}
where we expanded $U_{x\mu} \simeq 1+i\theta_{x\mu}$ by
assuming  small gauge-field fluctuations $\theta_{x\mu}$. 
This expression $M_G$ at a fixed $c_2$ certainly increases 
as $c_1\ (> c_{1c})$ increases.

%---------------------------------------------------
%----------------------------------------------------

%---------------------------------------------------
%----------------------------------------------------

An important point is that the gauge boson mass is 
{\em nonvanishing} on the critical line except for relatively 
large values of $c_2$.
To confirm this point we  present the size-dependence of $M_G$
in Figs.\ref{fig.pm4}-\ref{fig.pm6}.
They show that $M_G$ slightly increases  for a larger lattice,
which denies the possibility that $Min[M_G(c_1)] \rightarrow 0$ as 
$L \rightarrow \infty$.
This point is in contradiction to the expression (\ref{mftmass}),
where $\langle z_x \rangle^2 = 0$ at $c_1 = c_{1c}$ implies $M_G(c_{1c})=0$.
The nonperturbative effects of $U_{x\mu}$ are crucial to explain $M_G \neq 0$.
In fact, we  observe $M_G$ at the criticality increases as $c_2$ 
decreases.
This behavior is consistent with the result of the instanton 
density at $c_1=c_{1c}$, which is nonvanishing  and increases as $c_2$ 
decreases.
This suggests that on the critical line, 
the {\em confinement phase} in which instantons survive, rather than the Higgs
or Coulomb phases, is realized. In later subsection, we shall confirm this point
by studying the noncompact version of the present model.
Here we note that the fact $M_G(c_{1c}) \neq 0$ does not contradict
the general property of a second-order transition that 
the correlation length diverges at a critical point. 
In fact, we have observed in Fig.\ref{fig.ss2} that 
the spin gap $M_S$ vanishes at $c=c_{1c}$.
Thus the spin correlation length diverges on the critical line, which
is sufficient to generate singularities in thermodynamic quantities,
although the gauge correlation length is finite there.

%---------------------------------------------------
%----------------------------------------------------

%---------------------------------------------------
%----------------------------------------------------

%---------------------------------------------------
%----------------------------------------------------

%%%%%%%%%%%%%%%%%%%%%%%%%%%%%%%%%%%%%%%%%%%%%%%%%%%%%%%%%%%%%%%%%%

\subsection{Confinement phase vs Coulomb phase 
on the critical line}

Physical properties of a quantum system just on its critical point are 
interesting. For example in the previous paper\cite{YAIM}, by a 
gauge-theoretical study of a quantum spin
system having a N\`eel-dimer (Higgs-confinement) transition,
we argued that the Coulomb-like phase is 
realized on the critical point.
This is because there appears massless relativistic spinon $z_{xa}$ 
on the critical point and it
generates nonlocal correlations for the gauge field.
Typical form of the effective gauge model is given as 
\begin{equation}
S_{\rm eff}= g_e \sum_{\cal C} \gamma^{|{\cal C}|}\prod_{\cal C}U,
\label{Seff}
\end{equation}
where ${\cal C}$ denotes an arbitrary closed loop on the lattice
and $|{\cal C}|$ is its length.
$g_e$ is the effective gauge coupling constant and
$\gamma$ is a parameter of the hopping expansion.  
At the critical value $\gamma=\gamma_c$, which corresponds to 
relativistic massless matter fields,
$S_{\rm eff}$ diverges for the uniform configuration $U_{x,\mu}=1$.
Recently we studied  models closely related to Eq.(\ref{Seff}) and 
showed that there exists a deconfinement phase transition
at $\gamma=\gamma_c$ and sufficiently large $g_e$\cite{nonlocal}.
{\em In the deconfinement phase}, the 
renormalized gauge boson propagator behaves 
at long distance $p \sim 0$ as follows;
\begin{equation}
\langle A_\mu(-p)A_\nu(p)\rangle \sim 
 {1  \over \sqrt{p^2}}\ \delta_{\mu\nu},
\label{gp}
\end{equation}
which gives potential energy between charges at distance $r$ as
$1/r$.
Then we should evaluate the following Fourier integral instead of
Eq.(\ref{CG2}),
\begin{equation}
F(p^2_1+p^2_2;t)=\int dp_3{e^{ip_3t} \over \sqrt{\vec{p}^2+M^2}}.
\label{CFS2}
\end{equation}
We numerical evaluated the above integral (\ref{CFS2}) and found 
\begin{equation}
F(\ell;t) \sim e^{-\alpha_0 \sqrt{\ell+M^2}t}, \;\; \alpha_0 \sim 1.287...
\label{Fell}
\end{equation}

One may expect that the above change of the fitting function 
$F(\ell;t)$
may explain the finiteness of the gauge-boson mass at the 
criticality. However this is not the case.
Let us assume $M_G=0$ on the critical line in the present model. 
Then the correlation function of $\tilde{O}(x_3)$
obtained by the propagator (\ref{gp}), $D'_G(t)$,
behaves as 
\begin{equation}
D'_G(t)|_{M_G=0} \sim e^{-\alpha_0 \sqrt{p^2_1+p^2_2}\;t}.
\label{C'G}
\end{equation}
From Eq.(\ref{C'G}) it is obvious that the function 
$D'_G(t)|_{M_G=0}$
is {\em invariant} under a {\em scale transformation}; 
$(t, \vec{p})\rightarrow (\xi t,\vec{p}/\xi)$ where $\xi$ is a
patameter of the scale transformation.
This means that, if $M_G=0$ and the ``Coulomb phase" is realized
at the critical line, the damping factor of 
$D_G(t)$ stays {\em constant} along the critical line.
However in Fig.\ref{fig.pm2}, the value of $M_G$ at minimum
(which gives the damping factor) clearly increases as $c_2$ decreases.
This implies that there exists a physical scale of the mass dimension
(i.e. the gauge-boson mass $M_G$) in the gauge-boson sector at the
criticlity, which supports the conclusion in the previous section that
$M_G\neq 0$ on the critical line.

\begin{figure}[htbp]
\begin{center}
\leavevmode
\epsfxsize=8cm
\epsffile{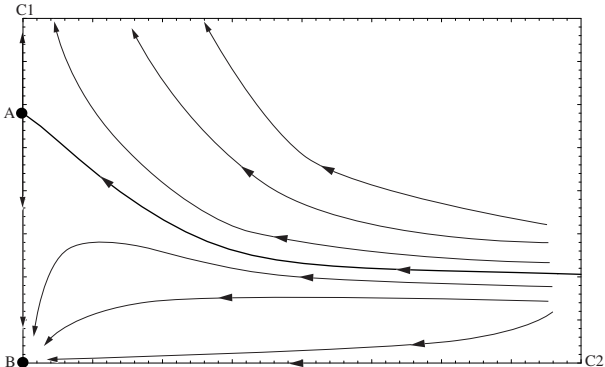}
\caption{Renormalization-group flow in the phase diagram.
The thick line is the critical line. A is an unstable fixed point,
and B is a stable fixed point where density of instantons
diverges.}
\label{fig.rg}
\end{center}
\end{figure}

In Fig.\ref{fig.rg} we present the suggested flow diagram of 
the renormalization group (RG) in the present model.
The RG flow on the {\em  critical line}
moves toward the unstable fixed point A in the phase diagram.
The two flow diagrams
 Fig.\ref{fig.rg} and Fig.\ref{fig.flow} are quite similar;
The unstable fixed point
($s=s_c, f=0$)  in Fig.\ref{fig.flow} corresponds to the point A
in Fig.\ref{fig.rg}, and similarly the N\'eel point to
the point $(c_2=0, c_1=\infty)$, the VBS to the point B($c_2=c_1=0$),
and the U(1) spin liquid to the point ($c_2= \infty, c_1=0$). 
More comments on this point will be given in Sec.8.

%---------------------------------------------------
%----------------------------------------------------

%%%%%%%%%%%%%%%%%%%%%%%%%%%%%%%%%%%%%%%%%%%%%%%%%%%%%%%%%%%%%%%%%%%%%%%%

\subsection{The case of noncompact U(1) gauge field}

For the U(1) gauge-Higgs model on a 3D lattice, exactly 
the same behavior of the
gauge-boson mass as ours was observed in the parameter region 
near the continuous phase transition\cite{GH1,AHM}.
In Ref.\cite{AHM}, by using the gauge fixing condition,
the gauge field propagator is investigated.
There the gauge field was splitted into a singular (instanton) 
part and the remaining regular part.
On the critical line and also in the confinement phase
it was shown that propagator of the regular part 
has a vanishing mass, whereas the full propagator has a finite mass.
From this observation, it is obvious that the finite mass at 
the criticality stems from the instanton gas. 

In order to check whether the above observation\cite{GH1,AHM}
on the generation of gauge-boson mass  on the critical line
can be applicable also to  the present model, one needs to
separate the effect of nonperturbative instanton effects.
In Ref.\cite{mv} the related work was reported for 
the O(3) sigma model, in which the effect of nonperturbative 
spin configuration was studied by suppressing the hedgehog configurations. 
Below we study this problem by considering the  
{\em noncompact} U(1) gauge theory of the CP$^1$ Schwinger bosons 
directly and compare the
results with those of the compact U(1) system.
The action of the noncompact model is given by
\begin{eqnarray}
S&=&-\frac{c_1}{2}\sum_{x,\mu,a}\Big(\bar{z}^a_{x+\mu}U_{x,\mu} 
z^a_x + \mbox{H.c.}\Big) \nonumber  \\
&&
+\frac{e_2}{2}\sum_x\sum_{\mu<\nu}(\Theta_{x,\mu\nu})^2
\label{model_2}
\end{eqnarray}
where $\Theta_{x,\mu\nu}$ is defined by Eq.(\ref{Theta})
and the constant $e_2$ controls fluctuations of the noncompact 
gauge field.
For small fluctuations of gauge field, $|\Theta_{x,\mu\nu}|\ll 1$, 
hence the two models are almost equivalent under  
the relation $e_2 \sim c_2$.

\begin{figure}[htbp]
\begin{center}
\leavevmode
\epsfxsize=8cm
\epsffile{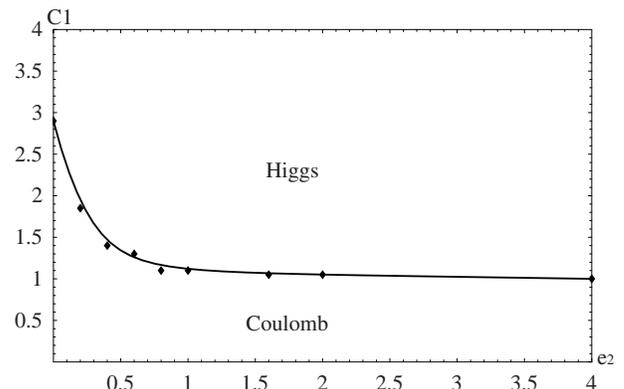}
\caption{Phase diagram of the noncompact gauge model.
In the noncompact case, the region of the ordered-Higgs phase 
is enlarged by the suppression of gauge-field fluctuations. 
The crossover line from the
dense to dilute instanton regions disappears.}
\label{fig.noncompact}
\end{center}
\end{figure}

In Fig.\ref{fig.noncompact} we present the phase diagram
of the model (\ref{model_2}) obtained by calculating the 
specific heat. The crossover line in the compact case disappears 
as we expect from the discussion on the instanton.
The region of the ordered-Higgs phase is enlarged by 
the suppression of fluctuations of $\theta_{x\mu}$.

%---------------------------------------------------

\begin{figure}[htbp]
\begin{center}
\leavevmode
\epsfxsize=8cm
\epsffile{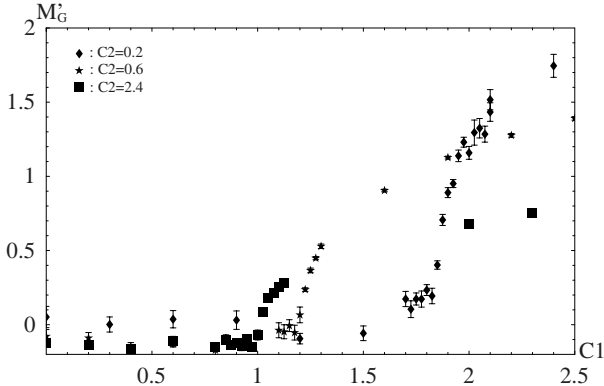}
\caption{Gauge-boson mass in the noncompact gauge model
as a function of $c_1$.
}
\label{fig.noncompactm1}
\end{center}
\end{figure}

In Fig.\ref{fig.noncompactm1} we present the gauge-boson mass 
$M'_{G}$ calculated as in the compact case.
It is obvious that  
$M'_G\sim 0$ in the disordered phase, and it starts
to increase at the critical point as expected.
Its behavior in the ordered-Higgs phase is more or less similar to
that of the compact gauge model.

\begin{figure}[htbp]
\begin{center}
\leavevmode
\epsfxsize=8cm
\epsffile{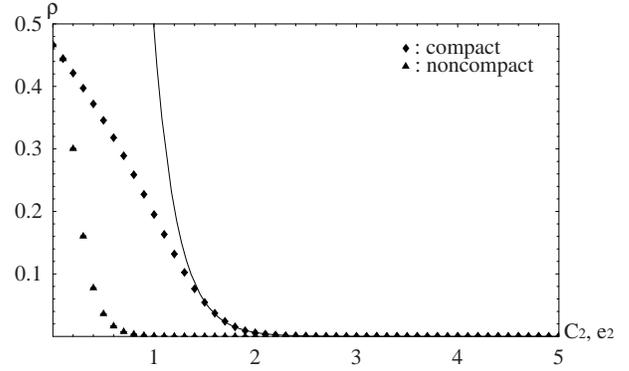}
\caption{Instanton density in compact and noncompact U(1)
gauge theories. The solid curve is a fit $\rho \propto \exp(-\ell c_2)$}
\label{fig.insnc}
\end{center}
\end{figure}

Then we are interested in the instanton density in the noncompact model.
From the definition of the instanton density (\ref{instden}),
it is obvious that instantons {\em can be generated} even in the 
noncompact model.
In Fig.\ref{fig.insnc} we compare the instanton density in the two
models, 
$\rho(c_2)$ in the compact model and 
$\rho'(e_2)$ in the noncompact model.

%---------------------------------------------------

%---------------------------------------------------
\begin{figure}[htbp]
\begin{center}
\leavevmode
\epsfxsize=8cm
\epsffile{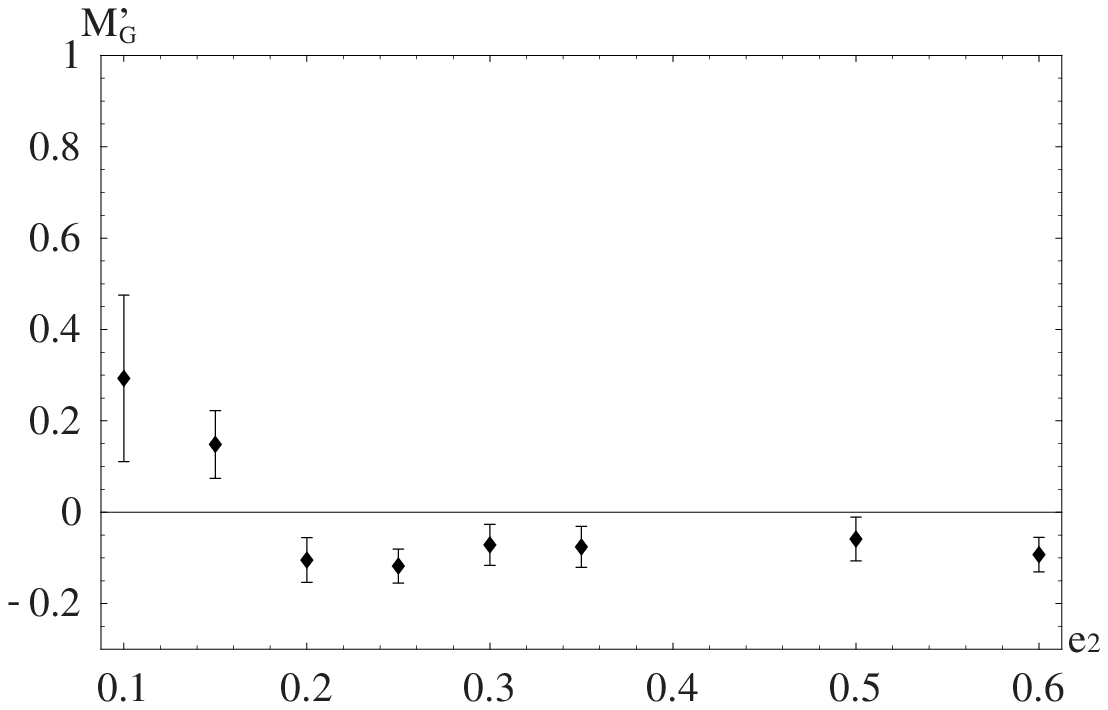}
\caption{Gauge-boson mass in the noncompact gauge model
as a function of $c_2$ with $c_1=0$.
}
\label{fig.noncompactm2}
\end{center}
\end{figure}

%---------------------------------------------------

From the results, it is easily seen that $\rho'(e_2)$ in the
noncompact gauge model is smaller than $\rho(c_2)$
in the compact model as expected.
However there exists a number of instantons for small $e_2$ 
(i.e., large gauge coupling) and one may wonder whether a
confinement phase exists for sufficiently small $e_2$.
The observation of the specific heat and the gauge-boson mass 
of course denies this possibility.

From the studies on the noncomapct gauge theory in this subsection,
it is obvious that the observed nonvanishing gauge-boson mass on the 
critical line of the comapct gauge theory stems from the existence of
instantons and not the Anderson-Higgs mechanism.

%%%%%%%%%%%%%%%%%%%%%%%%%%%%%%%%%%%%%%%%%%%%%%%%%%%%%%%%%%%%%%%%%%%%

\section{Discussion}
\setcounter{equation}{0}

In this paper, we have introduced 
 3D CP$^1$+U(1) lattice gauge theory and studied its phase
structure intensively. The specific heat exhibits
a second-order phase transition at $c_1 =c_{1c}(c_2)$.
The O(3) spin correlation functions show that this critical point
separates the spin-ordered and disordered phases.
The universality class changes from the O(3) spin model at
$c_2=0$ to the O(4) spin model $c_2 = \infty$ smoothly.
The MFT interprets the spin ordered phase as the Higgs phase and the
spin-disordered phase as the confinement phase. 
The distributions of instantons are studied, which 
provides us with a rough image of gauge-field configurations
in various points in the $c_2-c_1$ plane.
The gauge-invariant mass $M_G$ of gauge boson has two
origins; (i) condensation of instantons in the confinement phase
and (ii) Anderson-Higgs mechanism (condensation of the CP$^1$ field)
in the Higgs phase. As explained above, the conventional MFT 
plus one-loop correction like Eq.(\ref{mftmass})
predicts that $M_G=0$ on the critical points. However, we observed
$M_G \neq 0$ at $c_1= c_{1c}$ due to the remaining instantons.
This suggests that the system is in the confinement phase 
just on $c_1 = c_{1c}(c_2)$. $M_G(c_{1c}) \neq 0$ is compatible
with the second-order phase transition because the spin gap $M_S(c_{1c})=0$.

In our recent paper\cite{nonlocal}, we introduced  a 3D nonlocal U(1) 
lattice gauge theory, which is obtained  by mimicking the effect of
massless (gapless)  matter fields in a form
similar to  the gauge model Eq.(\ref{Seff}). 
This model contains an inverse gauge coupling $g_e$,
which is propotional to the number of massless matter fields.
By studying it numerically, we found that there exists a 
deconfinement phase transition at a critical 
gauge coupling $g_{e}=g_{ec}$.
This result suggests that a similar deconfinement phase may appear
on the critical points in the present model if the number of 
matter fields increases as CP$^1$ $\rightarrow$ CP$^N$ ($N$=large).
This possibility is also consistent with the recent analytical 
studies on the massless QED$_3$ which suggest that a 
deconfinement phase is realized for a sufficiently large number 
of massless fermions\cite{QED3}.

From these motivations, we numerically studied the CP$^N+$ 
U(1) gauge theory with $N=2,3$ and $4$, in particular its gauge-boson mass.
This model  has a qualitatively same phase structure as 
the CP$^1$+U(1) model but the critical line $c_1=c_{1c}(c_2)$ 
shifts upward in the phase diagram (shrinkage of the Higgs phase).

%---------------------------------------------------
\begin{figure}[htbp]
\begin{center}
\leavevmode
\epsfxsize=8cm
\epsffile{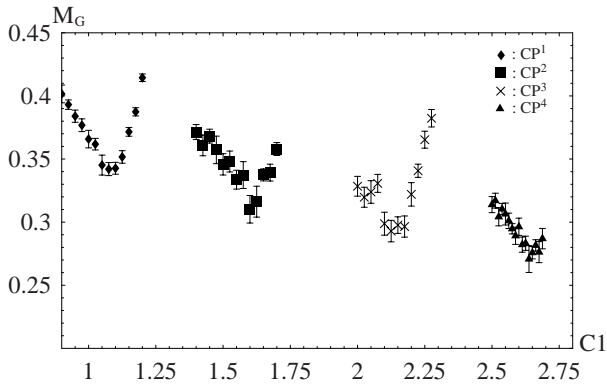}
\caption{Gauge-boson mass near the critical points in CP$^N$ ($N=1\sim 4$)
models.}
\label{fig.CP23}
\end{center}
\end{figure}
In Fig.\ref{fig.CP23}, we compare  the 
gauge-boson mass for $N=1,2,3$ and $4$.
$M_G$ decreases slightly as $N$ increases 
but is still nonvanishing.
We hope to report on more systematic study
of the CP$^N$+U(1) model in a future publication.

The present study should certainly shed some light 
on rich properties of wide range of  quantum spin systems
as a good Ginzburg-Landau-type phenomenological model.
For example, the observed phase diagram and the study of the instanton
in the present CP$^1$+U(1) model are quite useful to understand
the phase structure of the continuum GLT of Eq.(\ref{Ldec}).
As pointed out in Sec.7C, the RG flows in Fig.\ref{fig.flow} and
Fig.\ref{fig.rg} are very similar.
 Furthermore, we note that the instanton distributions in
Fig.\ref{fig.qbars} show that the most instantons
have topological charges $|Q_x|$ less than 3. For the AF Heisenberg
model with ring exchange, it was argued\cite{Bphase} that instantons with
$|Q_x|$ =1,2 and 3 do not contribute to disordering the
gauge dynamics because of a cancellation mechanism due to 
the Berry phase. Therefore, if there exist the  Berry phase term
in the action of the present model (\ref{model_1}), the Coulomb phase
is realized on the critical line instead of the confinement phase
and the GLT of Eq.(\ref{Ldec}) describes  this phase transition.\\

\vspace{1.5cm}

{\bf Acknowlegement}\\

We would like to thank Mr.T.Hiramatsu for discussion.

%%%%%%%%%%%%%%%%%%%%%%%%%%%%%%%%%%%%%%%%%%%%%%%%%%%%%%%%%%%%%%%%%%%%%%

\end{document}